\documentclass[referee,sn-mathphys,Numbered]{sn-jnl}

\usepackage{bm}
\usepackage{ulem}
\usepackage{lineno}
\usepackage{graphicx}
\usepackage{multirow}
\usepackage{amsmath,amssymb,amsfonts}
\usepackage{amsthm}
\usepackage{mathrsfs}
\usepackage[title]{appendix}
\usepackage{xcolor}
\usepackage{textcomp}
\usepackage{manyfoot}
\usepackage{booktabs}
\usepackage{algorithm}
\usepackage{algorithmicx}
\usepackage{algpseudocode}
\usepackage{listings}
\usepackage{array}
\usepackage{tabularray}
\usepackage{geometry}
\usepackage{graphicx}
\begin{document}

\title{Far-field Boundary Conditions for Airfoil Simulation at High Incidence in Steady, Incompressible, Two-dimensional Flow}

\author*[1]{\fnm{Narges} \sur{Golmirzaee}}\email{narges.golmirzaee@ucalgary.ca}

\author[1]{\fnm{David H.} \sur{Wood}}\email{dhwood@ucalgary.ca}

\affil[1]{\orgdiv{Department of Mechanical and Manufacturing Engineering}, \orgname{University of Calgary}, \orgaddress{
\city{Calgary}, \postcode{T2L 1Y6}, \state{Alberta}, \country{Canada}}}

\abstract{This study concerns the far-field boundary conditions (BCs) for airfoil simulations at high incidence where the lift and drag are comparable in magnitude and the moment is significant.  A NACA 0012 airfoil was simulated at high Reynolds number with the Spalart-Allmaras turbulence model in incompressible, steady flow.  We use the impulse form of the lift, drag, and moment equations applied to a control volume coincident with the square computational domain, to explore the BCs.  It is well known that consistency with the lift requires representing the airfoil by a point vortex, but it is largely unknown that consistency with the drag requires a point source as was first discovered by \citet{lagally1922berechnung} and \citet{filon1926}.  We show that having a point source in the BCs is more important at high drag than using a point vortex. The reason is that BCs without a point source cause blockage at the top and bottom sidewalls in a manner very similar to wind tunnel blockage for experiments.  A simple ``Lagally-Filon'' correction for small levels of blockage is derived and shown to bring the results much closer to those obtained using boundary conditions including a point source.  Although consistent with the lift and drag, the combined point vortex and source boundary condition is not consistent with the moment equation but the further correction for this inconsistency is shown to be very small.  We speculate that the correction may be more important in cases where the moment is critical, such as vertical-axis turbines.}


\maketitle
\section{Introduction}
\label{P4-sec:Intro}
%%%%%%%%%%%%%%%%%%
The  incompressible, two-dimensional, steady flow over a body like an airfoil at  high Reynolds number ($Re$),  is an important and fundamental problem in computational fluid dynamics (CFD). The accuracy of the computed lift, $L$, drag, $D$, and moment, $M$, is dependent on many factors, including but not limited to: the domain size and number of cells; the accuracy of the turbulence model; the quality of the numerical mesh; the differencing scheme; and the boundary conditions (BCs) \cite{versteeg}.  Here we consider the effect of the last-mentioned, motivated in part by the findings of \citet{roy2018} that accurate simulations of the flow around a NACA 0012 airfoil apparently requires a computational domain (CD) of size $A=500c$, where $A$ is the ``radius'' of the CD multiplied by the chord ($c$).  In practice, most domains are much smaller, typically around 30$c$, and it is obvious that domain size needs to be minimized to allow fast computation  for tasks like airfoil optimization.  Some of the desirable physical, numerical, and practical features of BCs  are described and investigated by \citet{choudhary2016}.  Here we take a different approach to investigate the following hypothesis:  BCs consistent with the equations for $L$, $D$, and $M$ for a control volume (CV) coincident with the computational domain, will minimize the domain size for a desired level of accuracy or increase the accuracy for a given domain size.

We previously studied the effects of BCs and domain size on the steady, incompressible, two-dimensional flow over a NACA 0012 airfoil at $Re = 6\times10^6$  for angles of attack, $\alpha$, of $5^{\circ}$, $10^{\circ}$, and $14^{\circ}$,  \citet{golmirzaee2024some}.  At these angles, the drag is small and has little effect  on the BCs, so we concentrated on the effect of the lift.  Following \citet{thomas1986} and \citet{destarac2011}, we found that the point vortex boundary condition (PVBC) ensures consistency\footnote{The terms ``consistency'' and ``consistent'' appear throughout this paper.  A full explanation of their meaning is delayed Section \ref{P4-sec:BC} to allow it to be stated precisely.} with the CV equation for $L$, the Kutta-Joukowsky (KJ) theorem, which was violated by the other four commonly-used BCs that we considered.   Of these, we concentrated on the set  BC-3 as it was the best performing of the four.  BC-3 is used here and described in Section \ref{P4-subsec:ConvBC}. By comparison to the results for a domain size of 500$c$, we showed the PVBC increased the accuracy of simulations for all smaller domains or allowed a reduction in domain size while maintaining a specific level of accuracy \cite{golmirzaee2024some}.  BCs that did not include a point vortex (PV), such as BC-3, increased the drag with the error depending on $C_{\mathrm{l}}^2c/A$, where $C_{\mathrm{l}}$ is the usual lift coefficient.  The drag coefficient, $C_{\mathrm{d}}$, was over-estimated by about $8\%$ when $A=30c$ at $\alpha = 14^\circ$, with the error occurring mainly in the pressure drag; 
errors of the same form were found in the Euler simulations of  \citet{destarac2011}.   $C_{\mathrm{l}}$ increased as $A$ increased, so that the aerodynamically-important $L/D$ ratio was doubly in error.  We did not study the moment because it is well known that for low angles, the center of pressure is very close to the quarter-chord position and the usual moment coefficient, $C_\mathrm{m}$, is close to zero.
%%%%%%%%%%%%%%%%%%%%%%%%%%
\begin{figure}
\centering
\includegraphics[width=0.8\linewidth]{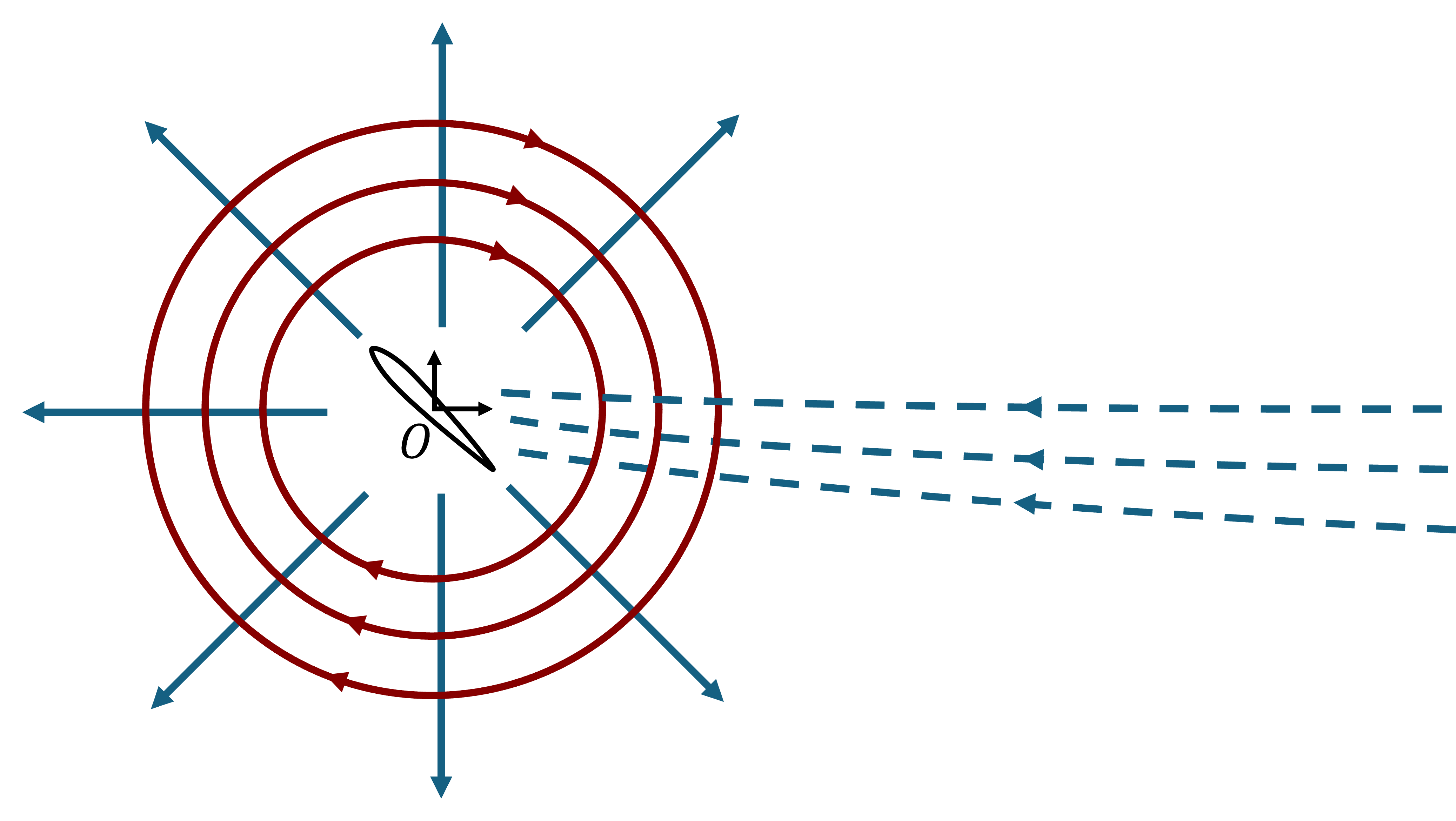}
\caption{The flow from left to right over an airfoil with a point vortex representing the lift and a point source compensating for the inflow in the wake. The wake is deflected in the direction opposite to the lift generated by the positive $\alpha$.  The figure is adapted from Fig. 5.12.4 of \cite{batchelor2000}.}
\label{P4-fig:PVPS-wake}
\end{figure}
%%%%%%%%%%%%%%%%%%%%%%%%%%

This study used the same NACA 0012 airfoil and $Re = 6\times10^6$, at a higher $\alpha=45^{\circ}$, where $D$ has the magnitude of $L$, and $M$ is also large. We show that the large $D$ makes the PVBC no more accurate than BC-3 and it is necessary to include a source term in the BCs, to produce a point-vortex-source-boundary condition (PVSBC).  The relationship between $D$ and source was derived first by \citet{lagally1922berechnung} and independently by \citet{filon1926}, see also \citet{liu2015}. A modern derivation is given in Section 5.12 of \citet{batchelor2000}, specifically his Eq. (5.12.15).

Modeling the drag as a point source (PS) causes the outward propagation of fluid in all directions to balance the inflow due to the velocity deficit in the wake, Fig. \ref{P4-fig:PVPS-wake}.   In contrast, BC-3 and PVBC prevent flow out of the top and bottom sidewalls of the square computational domain in Fig. \ref{P4-fig:Domain}. A partial correction for this sidewall blockage is derived and is shown to be very close to blockage corrections for  forces and moments measured in wind tunnels. 

We found only three previous uses of a PVSBC. \citet{kelmanson1987} used it for low-Re studies of flow over a circular cylinder. \citet{dannenhoffer1987grid}  applied it only for the drag associated with shock waves in compressible flow. \citet{allmaras2005} considered a PS and PV component to their BCs for incompressible airfoil flow at a low $\alpha=12^{\circ}$ where the drag is small. Somewhat confusingly, they cited \citet{dannenhoffer1987grid}  as the source for their source term. As far as we know, there has been no assessment of the effect of BCs on the accuracy of airfoil moment in the modern literature.  \citet{filon1928} derived an approximate solution for the far-field flow over a lifting body in laminar flow.  He found a logarithmic divergence in the moment equation for a CV in the far-field. \citet{goldstein1933} showed the divergence is related to the deflection of the wake in the direction opposite to the lift, but the full resolution of the problem had to wait until \citet{imai1951} provided a detailed higher order solution which canceled the logarithmic term.

The remainder of this paper has the following organization. Section \ref{P4-sec:ImpulseEq} presents the impulse equations for lift, drag, and moment. 
Section \ref{P4-sec:BC} describes the computational details and boundary conditions that are consistent with the lift and drag equations. It finishes by explaining why the point source BCs do not ensure consistency with the moment equation.  The results are presented and discussed in 
Section \ref{P4-sec:results}, followed by an assessment of the Imai correction in
Section \ref{P4-sec:ImaiCorr} which addresses the inconsistency with the moment equation; the correction is small for the present case.  
Section \ref{P4-sec:SummaryConclusion} provides a summary and the conclusions.

\section{Impulse equations for the lift, drag, and moment}
\label{P4-sec:ImpulseEq}
\begin{figure}
\centering
\includegraphics[width=0.55\linewidth]{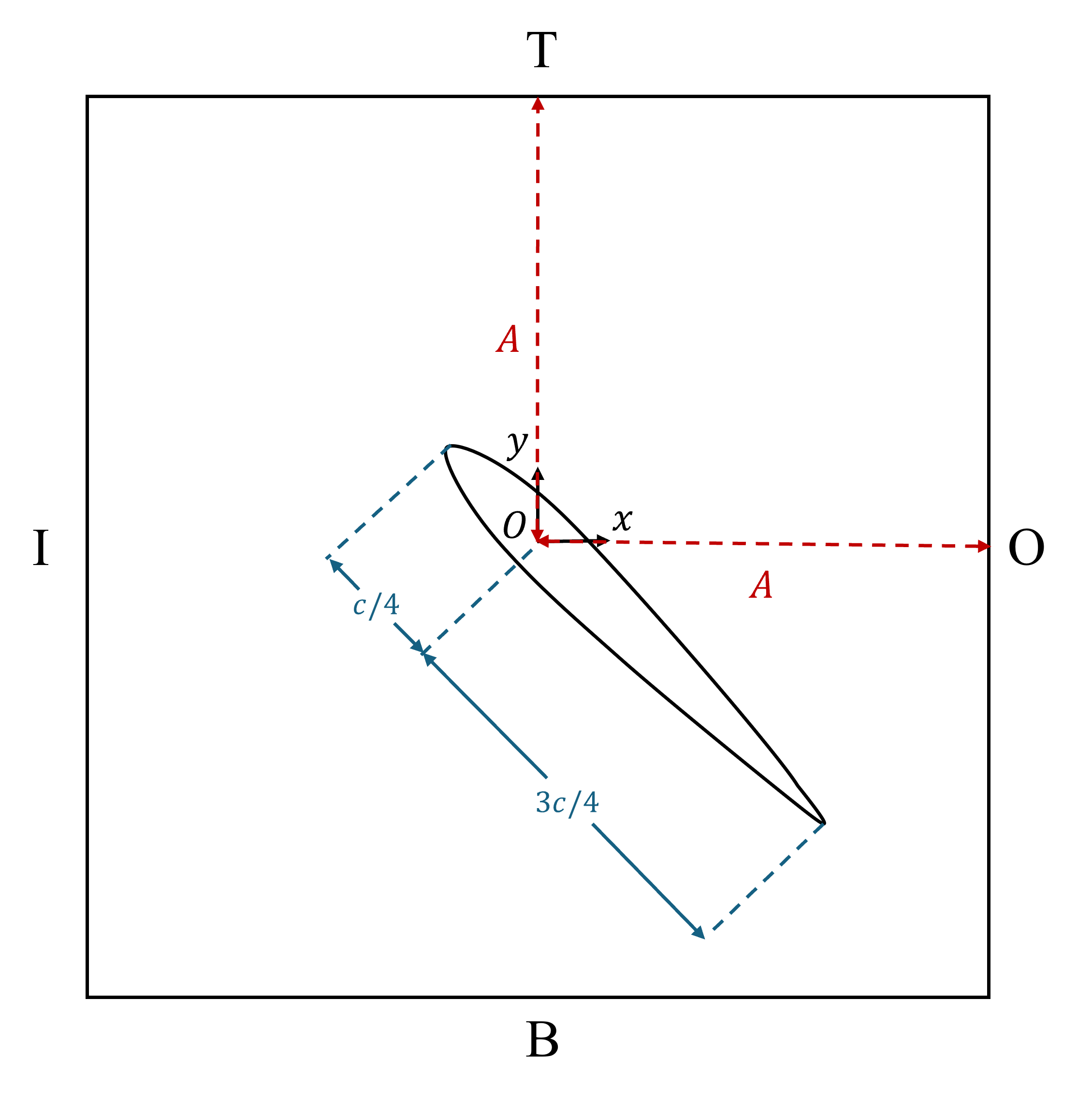}
\caption{The computational domain (CD) and control volume (CV) for the airfoil at $\alpha=45^{\circ}$. ``I'' indicates the inlet, ``T'' and ``B'' the top and bottom sidewalls, respectively, and ``O'' is the outlet.}
\label{P4-fig:Domain}
\end{figure}

All simulations used the square computational domain shown in Fig. \ref{P4-fig:Domain}. 
%%%%%%%%%%%%%%%%%%%%%%%%%%
%%%%%%%%%%%%%%%%%%%%%%%%%%
The point singularities were placed at the center of the airfoil whereas the origin of the Cartesian coordinates was at the quarter-chord to abide by convention. The boundaries are labeled as I (inlet), T (top), B (bottom), and O (outlet). Assuming that the wake containing all the vorticity that is not bound to the airfoil, exits from O only, the equations for the drag and lift for the CV coincident with this domain are reproduced from Eqs. (16) and (19) of \cite{golmirzaee2024some}.  They were derived from the Reynolds transport theorem forms of the ``impulse'' equations given by \citet{noca1997} who replaced the pressure in the conventional CV equations with the vorticity, $\Omega$. For the CV coincident with the domain boundaries in Fig. \ref{P4-fig:Domain}
%%%%%%%%%%%%%%%%%%%%%%%%%%
\begin{align}
\frac{D}{\rho}=&
\int_{I} 
\bigg(U_{\infty}u+\frac{1}{2}u^2-\frac{1}{2}v^2\bigg)\mathrm{d}y
%\nonumber \\ &
-\int_{T} 
(U_{\infty}v+uv)\mathrm{d}x
\nonumber \\ &
-\int_{O} 
\bigg[U_{\infty}u+\frac{1}{2}u^2-\frac{1}{2}v^2 + y(U_\infty+u)\Omega\bigg]\mathrm{d}y
%\nonumber \\ &
+\int_{B} 
(U_{\infty}v+uv)\mathrm{d}x
\label{P4-eq:Fx-impulse-Uinf-reduced}
\end{align}
%%%%%%%%%%%%%%%%%%%%%%%%%%
and
%%%%%%%%%%%%%%%%%%%%%%%%%%
\begin{align}
\frac{L}{\rho}=&
\int_{I} 
(U_{\infty}v +uv)\mathrm{d}y
+\int_{T} 
\bigg(U_{\infty}u+\frac{1}{2}u^2-\frac{1}{2}v^2\bigg)\mathrm{d}x
\nonumber \\ &
-\int_{O} 
(U_{\infty}v +uv)\mathrm{d}y
-\int_{B} 
\bigg(U_{\infty}u+\frac{1}{2}u^2-\frac{1}{2}v^2\bigg)\mathrm{d}x,
\label{P4-eq:Fy-impulse-Uinf-3}
\end{align}
%%%%%%%%%%%%%%%%%%%%%%%%%%
where $\rho$ is the density of air, and $u$ and $v$ are the perturbation velocities in the $x$ and $y$-directions from the free-stream velocity $(U_\infty,0)$. Using conservation of mass, we have
%%%%%%%%%%%%%%%%%%%%%%%%%%
\begin{equation}
U_{\infty}\bigg(\int_{I} 
u \mathrm{d}y
-\int_{T} 
v\mathrm{d}x
-\int_{O} 
u \mathrm{d}y
+\int_{B} 
v\mathrm{d}x \bigg)=0,
\label{P4-eq:Continuity-2}
\end{equation}
%%%%%%%%%%%%%%%%%%%%%%%%%%
so Eq. (\ref{P4-eq:Fx-impulse-Uinf-reduced}) becomes
%%%%%%%%%%%%%%%%%%%%%%%%%%
\begin{align}
\frac{D}{\rho}=&
\int_{I} 
\left(\frac{1}{2}u^2-\frac{1}{2}v^2\right)\mathrm{d}y
-\int_{T} 
uv\mathrm{d}x
\nonumber \\ &
-\int_{O} 
\left(\frac{1}{2}u^2-\frac{1}{2}v^2\right) \mathrm{d}y
-\int_{O}  y(U_{\infty}+u)\Omega\mathrm{d}y
+\int_{B} 
uv\mathrm{d}x.
\label{P4-eq:Fx-impulse-Uinf-reduced-2}
\end{align}
%%%%%%%%%%%%%%%%%%%%%%%%%%
The $U_{\infty}u$ and $U_{\infty}v$ terms in Eq. (\ref{P4-eq:Fy-impulse-Uinf-3}) give the circulation around the CV, Eq. (2) of \cite{golmirzaee2024some}, allowing Eq. (\ref{P4-eq:Fy-impulse-Uinf-3}) to be written as
%%%%%%%%%%%%%%%%%%%%%%%%%%
\begin{equation}
\frac{L}{\rho}= 
U_{\infty} \Gamma
+\int_{I} 
uv\mathrm{d}y
+\int_{T} 
\left(\frac{1}{2}u^2-\frac{1}{2}v^2\right)\mathrm{d}x
-\int_{O} 
uv\mathrm{d}y
-\int_{B} 
\left(\frac{1}{2}u^2-\frac{1}{2}v^2\right)\mathrm{d}x.
\label{P4-eq:Fy-impulse-Uinf-4}
\end{equation}
%%%%%%%%%%%%%%%%%%%%%%%%%%

The  equation for $M$ can be easily derived from Eq. (A.2.7) of \cite{siala2019} as
%%%%%%%%%%%%%%%%%%%%%%%%%%
\begin{align}
\frac{M}{\rho} = &
\int_{I}
\bigg(
uvx+U_{\infty}vx+\frac{1}{2}v^2 y -\frac{1}{2} U_{\infty}^2y-U_{\infty}uy-\frac{1}{2}u^2y \bigg) \mathrm{d}y
\nonumber \\
+ & 
\int_{T}
\bigg(
uvy+U_{\infty}vy
-\frac{1}{2}v^2 x
+\frac{1}{2} U_{\infty}^2x
+U_{\infty}ux +\frac{1}{2}u^2x \bigg) \mathrm{d}x
\nonumber \\
+ &
\int_{O}
\bigg(
\frac{1}{2}(x^2+y^2)(U_{\infty}+u)\Omega
-uvx-U_{\infty}vx-\frac{1}{2}v^2 y +\frac{1}{2} U_{\infty}^2y+U_{\infty}uy+\frac{1}{2}u^2y \bigg) \mathrm{d}y
\nonumber \\
+ & 
\int_{B}
\bigg(
uvy-U_{\infty}vy
+\frac{1}{2}v^2 x
-\frac{1}{2} U_{\infty}^2x
-U_{\infty}ux -\frac{1}{2}u^2x \bigg) \mathrm{d}x. \label{P4-eq_mom1}
\end{align}
%%%%%%%%%%%%%%%%%%%%%%%%%%
All three equations contain quadratic terms in $u^2, v^2$, and $uv$ which will become negligible compared to the linear terms like $U_\infty u$ if $A$ is sufficiently large.  Then, the force and moment coefficients are given by
%%%%%%%%%%%%%%%%%%%%%%%%%%
\begin{align}
C_{\mathrm{d}} =&
-\frac{2}{U_{\infty}c}\int_{O}  y \Omega\mathrm{d}y,
\label{P4-cd}
\end{align}
%%%%%%%%%%%%%%%%%%%%%%%%%%
\begin{align}
C_{\mathrm{l}} = 2\Gamma/(U_{\infty}c) &
\label{P4-cl}
\end{align}
%%%%%%%%%%%%%%%%%%%%%%%%%%
which includes the KJ equation, and
%%%%%%%%%%%%%%%%%%%%%%%%%%
\begin{align}\label{P4-cm1}
C_{\mathrm{m}} = &
\frac{2}{U_\infty c^2} \bigg(-\int_{I}
uy \mathrm{d}y
+\int_{O}
uy \mathrm{d}y
+\int_{T} vy \mathrm{d}x
-\int_{B} vy \mathrm{d}x 
\nonumber\\
 & +\int_{I} vx \mathrm{d}y
 -\int_{O} vx \mathrm{d}y
 +\int_{T} ux \mathrm{d}x -\int_{B} ux \mathrm{d}x\bigg) \nonumber \\ 
 &+\frac{1}{U_\infty c^2}\int_{O}y^2 \Omega  \mathrm{d}y.
\end{align}
%%%%%%%%%%%%%%%%%%%%%%%%%%
The RHS of Eq. (\ref{P4-cm1}) has to be multiplied by $-1$ to reproduce the conventional definition of aerodynamic moment so this will be done throughout the paper.  Further reduction of the vorticity integrals will be made below.

 Equation (\ref{P4-cm1}) is grouped into separate terms, one per each of the three lines, on the following basis.  The equation gives the moment about $(0,0)$ which, following convention, is the quarter-chord.  The moment about any $(x_1,y_1)$ must be given by
 %%%%%%%%%%%%%%%%%%%%%%%%%%
\begin{equation}
C_{\mathrm{m}}(x_1,y_1) = C_{\mathrm{m}} +  C_{\mathrm{l}}x_1/c-C_{\mathrm{d}}y_1/c.
\label{P4-cmx}
\end{equation}
%%%%%%%%%%%%%%%%%%%%%%%%%%
The first line of Eq. (\ref{P4-cm1})  contributes to $C_{\mathrm{m}}$, but to no other term in Eq. (\ref{P4-cmx}) by conservation of mass, the second line with the KJ equation gives the second term in Eq. (\ref{P4-cmx}), and the third line provides the third term.

Equation (\ref{P4-cd}) requires the integral of $\Omega y$ at the outlet be invariant as $A$ changes, and Eq. (\ref{P4-cl}) requires the same for the boundary integrals summing to $\Gamma$.  There is no similar constraint, however, on the  sum of the boundary integrals or the outlet integral of vorticity  in Eq. (\ref{P4-cm1}) as it is only their sum that is invariant. We will show below that this prevents the complete consistency of the BCs considered here with the moment equation.

There is an important consequence of the invariance of the circulation and the drag: the quadratic terms in Eqs. (\ref{P4-eq:Fx-impulse-Uinf-reduced-2}) and (\ref{P4-eq:Fy-impulse-Uinf-4}) must sum to zero for \textit{any} $A$.  To show this, take two domains, $A_1$ and $A_2$ with $A_2 \gg A_1$ such that the quadratic terms are negligible for $A_2$.  Since no circulation arises in the region outside of $A_1$ and the vorticity integral for the drag leaving $A_1$ will equal to the vorticity integral leaving $A_2$, the lift and drag quadratic terms must sum to zero for $A_1$.  This result holds for a point source and/or vortex located anywhere within $A$, for any value of $A$.  The same is true of the quadratic terms in Eq. (\ref{P4-eq_mom1}).
%%%%%%%%%%%%%%%%%%%%%%%%%%
\section{Computational details and boundary conditions}
\label{P4-sec:BC}
Many of the details of the present simulations are the same as for \cite{golmirzaee2024some},  so only the main differences are given here.  
We regard our simulations as having avoided or addressed most of the sources of error listed in the Introduction, and offer the following to support this contention. \citet{roy2018} described a code verification study of the NACA 0012 at $\alpha = 10^{\circ}$ and $Re= 6\times10^6$, using the Spalart-Allmaras turbulence model as used here and in \cite{golmirzaee2024some}.  \citet{roy2018} gave ``estimated upper and lower bounds on the benchmark solution'' as: 
$C_{\mathrm{l}}$, $1.0909$ – $1.0911$;  $C_{\mathrm{d}}$, $0.012265$ – $0.012275$, and; $C_{\mathrm{dp}}$, $0.00606$ – $0.00607$,
where   $C_{\mathrm{dp}}$ is the pressure drag coefficient.   The variation in the results shown in their Fig. 4 appear to exceed those bounds. \citet{golmirzaee2024some} obtained $C_{\mathrm{l}} = 1.0765$, $C_{\mathrm{d}} = 0.01219$, and $C_{\mathrm{dp}}=0.00590$ for $A=500c$.
We could not find any code verification study for the NACA 0012 at a higher angle of attack.

Section \ref{P4-subsec:OpenFMsetting} presents the simulation settings.
Section \ref{P4-subsec:ConvBC} discusses common BCs used in airfoil simulations, focusing on two specific ones, BC-3 and PVBC, that were considered by \citet{golmirzaee2024some}, and used here. In Section \ref{P4-subsec:PVSBC}, we introduce the point source addition to PVBC, which increases simulation accuracy.
%%%%%%%%%%%%%%%%%%%%%%%%%%
\subsection{Computational settings}
\label{P4-subsec:OpenFMsetting}
We utilized OpenFOAM with the simpleFOAM algorithm, which is well-suited for incompressible, steady, turbulent flows. The  solvers were as follows: for $p$, the GAMG solver with DICGaussSeidel smoother, and for $U$ and nuTilda, smoothSolver and DILUGaussSeidel as the solver and smoother, respectively. Table \ref{P4-table:schemes}  shows the different numerical schemes.

%%%%%%%%%%%%%%%%%%%%%%%%%%
\begin{table}
\centering
\caption{Different schemes used in the simulation settings of this study}
\begin{tabular}{ll}
\hline
Schemes & settings \\
\hline
ddtSchemes & steadyState \\
gradSchemes & Gauss linear \\
divSchemes &  \\
\qquad div(phi,U) & bounded Gauss linearUpwind grad(U) \\
\qquad div(phi,nuTilda) & bounded Gauss linearUpwind grad(nuTilda) \\
\qquad div((nuEff*dev2(T(grad(U))))) & Gauss linear \\
laplacianSchemes & Gauss linear corrected \\
interpolationSchemes & linear \\
snGradSchemes & corrected \\
\hline
\end{tabular}
\label{P4-table:schemes}
\end{table}
%%%%%%%%%%%%%%%%%%%%%%%%%%

The computational domain, depicted in Fig. \ref{P4-fig:Domain}, is a square of sides $2A$, with \text{$10c \le A \le 500c$}.
The aerodynamic center of the airfoil, located at the quarter-chord is the origin of the coordinates and the point about which the  moment was calculated.
As will be discussed in Section \ref{P4-subsec:Grids}, we ensured the creation of high-quality grids, allowing a high relaxation factor of 0.95. This approach was chosen to balance accuracy and  minimization of computational time.

%%%%%%%%%%%%%%%%%%%%%%%%%%
\subsection{Grids}
\label{P4-subsec:Grids}
\begin{figure}
\centering
\includegraphics[width=0.8\linewidth]{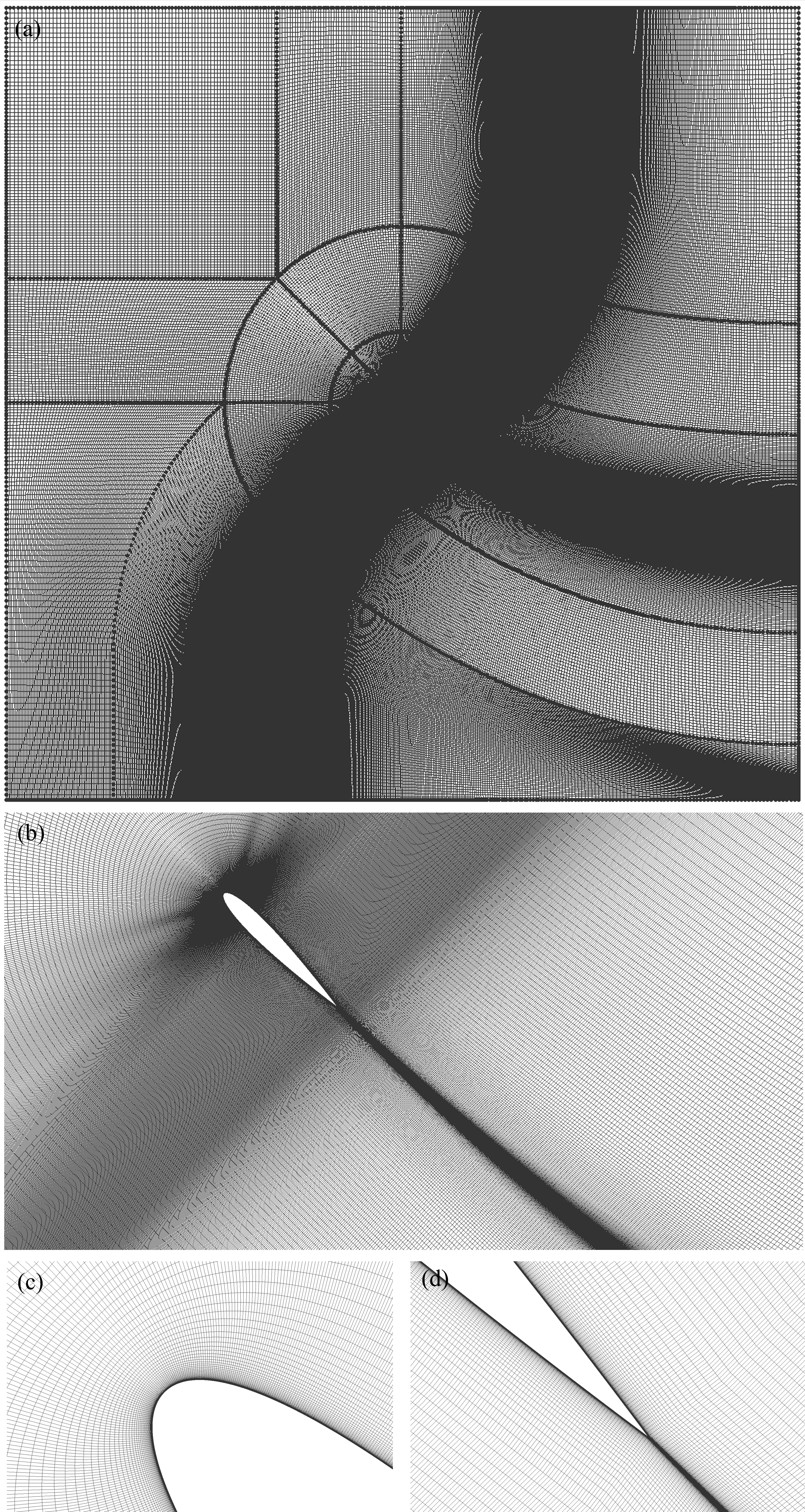}
\caption{Views of the intermediate computational mesh. (a) The whole mesh for $A=30c$. (b) More detail near the airfoil. (c) The mesh near the leading edge. (d) The mesh near the trailing edge.}
\label{P4-fig:Grids}
\end{figure}

Grids, like that shown in Fig. \ref{P4-fig:Grids}, generated using Pointwise, involved creating four layers of varying coarseness. Each layer contained four times the number of cells as the previous one, with approximately every other point of the finer grid omitted for the next coarser grid. Special attention was given to maintaining a smooth transition between adjacent cells to prevent jumps in the area ratio, which could otherwise induce spurious vorticity. Across all grids, the area ratio was kept below 1.19.
Also, we ensured that the wake  contained sufficiently refined cells with aspect ratios not exceeding $20$. The grids were designed to be perpendicular to avoid potential numerical errors at the boundaries. The first grid heights and growth ratios around the airfoil were aligned with those used in \cite{golmirzaee2024some}, resulting in a fine grid comprising $2,551,808$ cells for $A=30c$. The $y^+$ value across all points at the airfoil surface was kept below $0.1$, as recommended in \cite{ecca2018viscous}.
We also undertook similar studies of grid convergence as in \cite{golmirzaee2024some} to  ensure a low level of numerical uncertainty.
Using Roache’s \cite{roache1994perspective} method, we derived estimated values of $0.9082$ and $0.9207$ for $C_{\mathrm{d}}$ and $C_{\mathrm{l}}$, respectively, at zero grid spacing, with numerical uncertainty bounds for $C_{\mathrm{d}}$ and $C_{\mathrm{l}}$ of $0.002\%$ and $0.04\%$, respectively. The same grids were used for the three BCs studied.
%%%%%%%%%%%%%%%%%%%%%%%%%%
%%%%%%%%%%%%%%%%%%%%%%%%%%
\subsection{Boundary conditions}
\label{P4-subsec:ConvBC}
\citet{golmirzaee2024some} tested  four  BCs commonly used for airfoil simulations, which were recommended by the sources given in \cite{golmirzaee2024some}.  They apply either, fixed, freestream, slip, or symmetry conditions for the flow variables at the far-field boundaries.
\citet{golmirzaee2024some} showed the set designated BC-3 produced better or equivalent accuracy for the force coefficients compared to the other BCs. Moreover, BC-3 does not generate spurious vorticities at  T and B. Therefore, we employ BC-3 as a representative of the common boundary conditions.
\citet{golmirzaee2024some} also used the PVBC to ensure consistency with the KJ theorem. The induced velocities $u$ and $v$ at an arbitrary point $(x,y)$ are easily derived from the Biot-Savart law.  In doing so, we ignored the difference in the location of the PV and the origin of the coordinates,
so the PVBC for the velocity vector is
%%%%%%%%%%%%%%%%%%%%%%%%%%
\begin{equation}
(U_\mathrm{\infty}+u, v)=\left(U_{\infty}+\frac{ yU_{\infty}C_{\mathrm{l}} c}{4\pi (x^2+y^2)}, -\frac{xU_{\infty}C_{\mathrm{l}} c}{4\pi (x^2+y^2)}\right)\cdot
\label{P4-eq:PV-velocity}
\end{equation}
%%%%%%%%%%%%%%%%%%%%%%%%%%
For the pressure at the outlet boundary, Bernoulli's equation is applied and second order terms are ignored, leading to
%%%%%%%%%%%%%%%%%%%%%%%%%%
\begin{equation}
p=-\frac{\rho y U_{\infty}^2 C_{\mathrm{l}} c}{4 \pi (x^2+y^2)}\cdot
\label{P4-eq:PV-pressure}
\end{equation}
%%%%%%%%%%%%%%%%%%%%%%%%%%
Clearly, the application of PVBC depends upon the solution of the simulation. Every user-input number of iterations (typically 100),  a user-defined function updates the boundary pressure and velocities  using the lift calculated at that step. This iterative approach ensures that the boundary conditions remain consistent with the evolving flow field.
Table \ref{P4-table:BCs-SA} summarizes the values used for the flow variables at the computational domain boundaries for both BC-3 and PVBC.

We can now explain the term ``consistency'' that appears throughout the paper.  Ideal BCs must satisfy  Eqs. (\ref{P4-cd} -- \ref{P4-cm1}) exactly in the far-field when the quadratic terms are negligible, but this requires the outlet flow to also satisfy the BCs, and we have no direct control over this.  Further, it will be shown in Section \ref{P4-subsec:ForceMomentA500} that the quadratic terms in the $D$ and $M$ equations last for much larger values of $A$ than in the $L$ equation.
For these reasons we allow some latitude to the BCs and ask only that they be consistent with the CV equations by satisfying them as $A \to \infty$.
%%%%%%%%%%%%%%%%%%%%%%%%%%
\begin{table}
\centering
\caption{BC-3 and the point vortex boundary condition (PVBC) from \cite{golmirzaee2024some}. The airfoil BC for PVBC is the same as for BC-3.}
\begin{tabular}{lcccc}
\hline
& \multicolumn{4}{c}{BC-3}\\
\cmidrule{2-5}
Boundaries & $\bm{U} \ [ms^{-1}]$ & $p \ [m^2s^{-2}]$ & $\nu_t \ [m^2 s^{-1}]$  & $\tilde{\nu} \ [m^2 s^{-1}]$ \\
\hline
I & fixedValue, & zeroGradient & fixedValue, & fixedValue, \\
& $(51.48, 0)$ &  & $8.58 \times 10^{-6}$  & $3.432 \times 10^{-5}$ \\
T, B & slip & slip & slip  & slip \\
O & zeroGradient & fixedValue, $0$ & zeroGradient & zeroGradient \\
Airfoil & fixedValue, & zeroGradient & fixedValue,  & fixedValue,  \\
 & $(0, 0, 0)$ &  &  $0$ &  $0$ \\
\hline \\
 & \multicolumn{4}{c}{PVBC}\\
\cmidrule{2-5}
Boundaries & $\bm{U} \ [ms^{-1}]$ & $p \ [m^2s^{-2}]$ & $\nu_t \ [m^2 s^{-1}]$  & $\tilde{\nu} \ [m^2 s^{-1}]$ \\
\hline
I, T, B & Eq. (\ref{P4-eq:PV-velocity}) & zeroGradient & fixedValue,  & fixedValue, \\
 &  &  &  $8.58 \times 10^{-6}$  &  $3.432 \times 10^{-5}$ \\
O & zeroGradient & Eq. (\ref{P4-eq:PV-pressure}) &  zeroGradient & zeroGradient \\
\hline
\end{tabular}
\label{P4-table:BCs-SA}
\end{table}
%%%%%%%%%%%%%%%%%%%%%%%%%%
\subsection{Adding a point  source to the boundary conditions}
\label{P4-subsec:PVSBC}
The need for a PS in the BCs follows from consideration of the left side of Eq. (\ref{P4-eq:Continuity-2}). If we make a Helmholtz decomposition of $(u,v)$ into  components due to the vorticity, with subscript $\mathrm{v}$, and the remaining irrotational components, which need not be subscripted, we have
%%%%%%%%%%%%%%%%%%%%%%%%%%
\begin{equation}
\int_{I} 
u \mathrm{d}y
-\int_{T} 
v\mathrm{d}x
-\int_{O} 
u \mathrm{d}y
-\int_{O} 
u_{\mathrm{v}} \mathrm{d}y
+\int_{B} 
v\mathrm{d}x =0
\label{P4-eq:Continuity-3}
\end{equation}
%%%%%%%%%%%%%%%%%%%%%%%%%%
with the integral for $u_{\mathrm{v}}$ confined to the wake.  For sufficiently large $A$, the vorticity integral in Eq. (\ref{P4-cd}) for $D$ becomes
%%%%%%%%%%%%%%%%%%%%%%%%%%
\begin{align}
\int_{O}  y \Omega\mathrm{d}y = \int_{O} u_{\mathrm{v}}\mathrm{d}y,
\label{P4-cd1a}
\end{align}
%%%%%%%%%%%%%%%%%%%%%%%%%%
which is the far-field version of the conventional  equation for $D/(\rho U_\infty)$ in any wake, e.g., Eq. (2.84) of \citet{anderson2017fundamentals}. Now, the irrotational integrals in Eq. (\ref{P4-eq:Continuity-3}) are zero for BC-3 and PVBC for large values of $A$, so we have the inconsistent result that the drag must also be zero. Thus we need to add a PS to make the irrotational integrals non-zero.  By Eq. (\ref{P4-cd1a}), the strength of the source, $\Lambda$, is $D/(\rho U_\infty)$, as derived by \citet{lagally1922berechnung} and \citet{filon1926} and given by Eq. (5.12.15) of \cite{batchelor2000}. It is the drag equivalent of the KJ equation.  Given that Lagally and Filon were active only shortly after Kutta and Joukowsky, it seems remarkable that their equation, $D = \rho U_\infty \Lambda$, is so little known. There is a possible reason for this neglect: the relation between circulation and lift carries over to three dimensions, but the Lagally-Filon (LF) equation does not in general. For example, in the Trefftz plane behind a finite wing, the induced drag must be related to the streamwise vorticity over the plane.  This vorticity, however, does not lead to a term in the  conservation of mass equation.  In other words, the three dimensional form of the LF equation involves only the drag associated with the vorticity lying in the Trefftz plane. A further consideration is that Lagally was not well-known outside Germany and the two papers of Filon cited here are the only ones he wrote on aerodynamics.
Returning to two-dimensions, the PVSBC for the velocity perturbations is
%%%%%%%%%%%%%%%%%%%%%%%%%%
\begin{equation}
(u, v)=\left(\frac{ yU_{\infty}C_{\mathrm{l}} c}{4\pi (x^2+y^2)}
+\frac{ xU_{\infty}C_{\mathrm{d}} c}{4\pi (x^2+y^2)}, 
-\frac{xU_{\infty}C_{\mathrm{l}} c}{4\pi (x^2+y^2)}+
\frac{yU_{\infty}C_{\mathrm{d}} c}{4\pi (x^2+y^2)}
\right),
\label{P4-eq:PVPS-velocity}
\end{equation}
%%%%%%%%%%%%%%%%%%%%%%%%%%
and using Bernoulli's equation, 
%%%%%%%%%%%%%%%%%%%%%%%%%%
\begin{equation}
p=-\frac{\rho y U_{\infty}^2 C_{\mathrm{l}} c}{4 \pi (x^2+y^2)}
-\frac{\rho x U_{\infty}^2 C_{\mathrm{d}} c}{4 \pi (x^2+y^2)}
\label{P4-eq:pvps-Bernoulli}
\end{equation}
%%%%%%%%%%%%%%%%%%%%%%%%%%
\begin{table}
\centering
\caption{Point vortex and source boundary condition (PVSBC). The CV and the airfoil BC were the same as in Table \ref{P4-table:BCs-SA}.}
\begin{tabular}{lcccc}
\hline
Boundaries & $\bm{U} \ [ms^{-1}]$ & $p \ [m^2s^{-2}]$ & $\nu_t \ [m^2 s^{-1}]$  & $\tilde{\nu} \ [m^2 s^{-1}]$ \\
\hline
I, T, B & Eq. (\ref{P4-eq:PVPS-velocity}) & zeroGradient & fixedValue,  & fixedValue, \\
 &  &  &  $8.58 \times 10^{-6}$  &  $3.432 \times 10^{-5}$ \\
O & zeroGradient & Eq. (\ref{P4-eq:pvps-Bernoulli}) &  zeroGradient & zeroGradient \\
\hline
\end{tabular}
\label{P4-table:PSBC}
\end{table}
%%%%%%%%%%%%%%%%%%%%%%%%%%

The PVSBC is, therefore consistent with Eq. (\ref{P4-cl}) for $L$, and Eq. (\ref{P4-cd}) for $D$, but not necessarily with Eq. (\ref{P4-cm1}) for $M$.  To show this, we note that the first two lines of Eq. (\ref{P4-cm1}) sum to zero for the PV and PS placed anywhere within the CV. Parenthetically, it is for this reason we did not seriously experiment with moving  the PS around the domain.  

The vorticity integral in Eq. (\ref{P4-cm1}) can be written as
%%%%%%%%%%%%%%%%%%%%%%%%%%
\begin{align}
\int_{O}  y^2 \Omega\mathrm{d}y = 2 \int_{O} y u_{\mathrm{v}}\mathrm{d}y.
\label{P4-cd2}
\end{align}
%%%%%%%%%%%%%%%%%%%%%%%%%%
We now make two simple assumptions about the wake:
%%%%%%%%%%%%%%%%%%%%%%%%%%
\begin{enumerate}
    \item{The velocity remains approximately symmetric about $y_\mathrm{m}$, the point of minimum $u_{\mathrm{v}}$.  The vorticity integral in Eq. (\ref{P4-cd2}) is then equal to $ -2y_\mathrm{m} D/(\rho U_{\infty})$, and}
    \item{The wake is deflected away from the lift by the circulation. The vertical velocity is approximately $-\Gamma/(2\pi x)$  so that $y_\mathrm{m} \sim -L\ln(x/c)/\rho $.}
\end{enumerate}
%%%%%%%%%%%%%%%%%%%%%%%%%%
Thus, the vorticity integral diverges logarithmically with increasing $A$, while the boundary integrals in the first two lines of Eq. (\ref{P4-cm1})  sum to zero for all $A$.  The rate of divergence is proportional to $C_{\mathrm{l}} C_{\mathrm{d}}$.   This is Goldstein's \cite{goldstein1933} simple explanation for the logarithmic divergence of the moment equation found by \citet{filon1928} after determining the next highest order terms in the far-field streamfunction after those for the PV and PS.  Ref.\cite{goldstein1933} ascribed the divergence to the deflection of the wake, as  $y_{\mathrm{m}}$ and the vorticity integral are both zero for a symmetric, non-lifting body. The divergence was finally removed by the even higher order analysis of \citet{imai1951}. We will consider Imai's analysis in 
Section \ref{P4-sec:ImaiCorr} after describing the accuracy of $L$, $D$, and $M$ for different values of $A$ and the BCs. We note that $C_{\mathrm{l}} C_{\mathrm{d}}$ is maximized for airfoils when $\alpha \approx 60^\circ$ so our choice of $\alpha = 45^\circ$, made before Goldstein's explanation was known to us, should make obvious any problems with consistency of the PVSBC with the moment equation.
%%%%%%%%%%%%%%%%%%%%%%%%%%
\section{Results and discussion}
\label{P4-sec:results}
%%%%%%%%%%%%%%%%%%%%%%%%%%
\subsection{Effect of domain size and boundary conditions on the lift, drag, moment, and execution time}
\label{P4-subsec:DiffBCsTimes}
It will become apparent that the generation of vorticity in the supposed irrotational flow outside the wake is an important issue in these simulations.  Spurious vorticity may be caused by imposing the Neumann boundary condition at O, which approximates $\Omega$ as $-\partial u/\partial y$ only and so possibly makes it non-zero. We examined whether velocity gradients deduced from PVSBC in the $x$-direction and applied at O give higher accuracy compared with using the Neumann condition of zero gradient.
For $A=30c$, including  $\partial u/\partial x$ and $\partial v/\partial x$ from Eq. (\ref{P4-eq:PVPS-velocity}), added to the execution time without altering $L$, $D$, or $M$.

\begin{figure}
\centering
\includegraphics[width=0.75\linewidth]{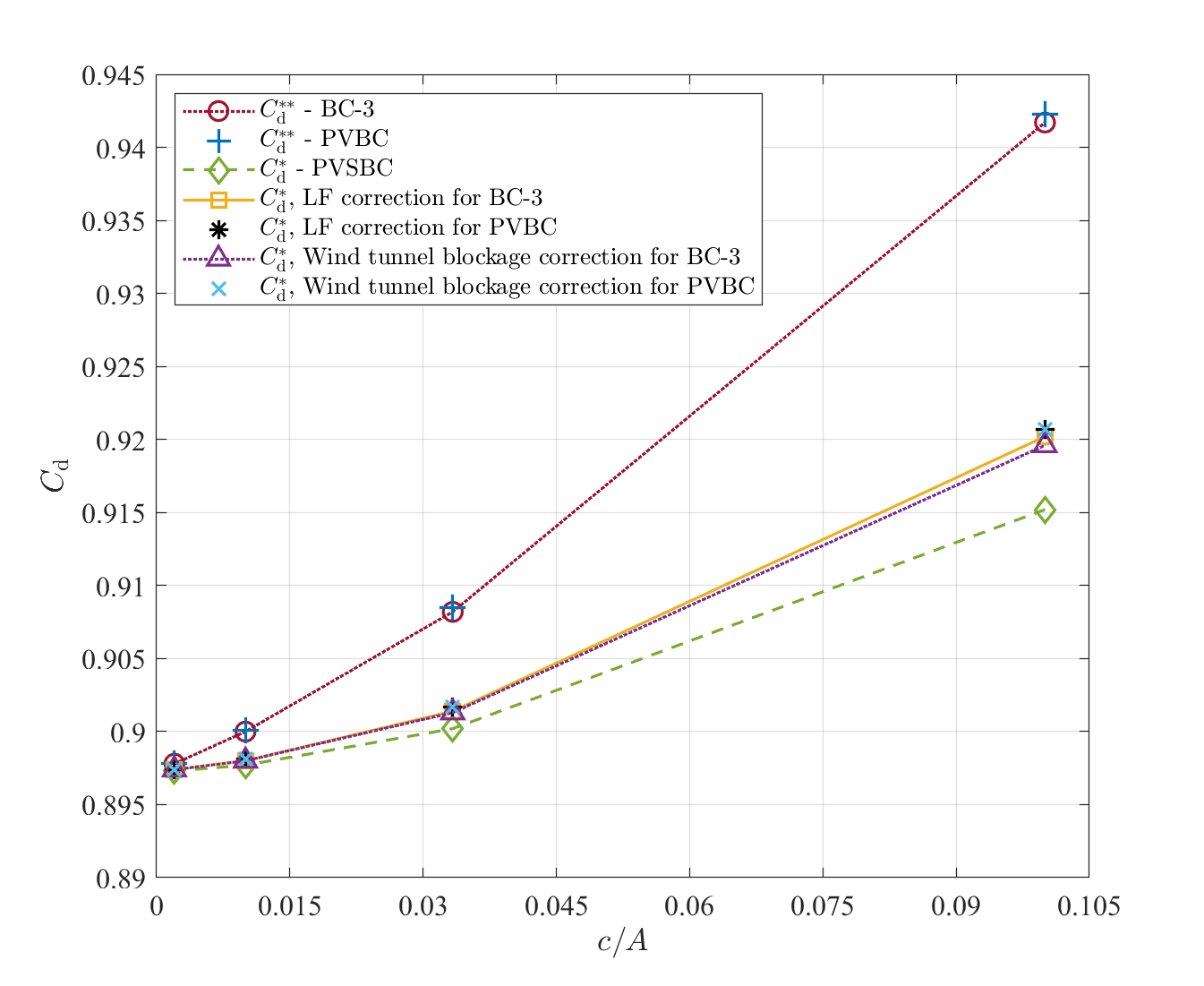}
\caption{Comparison of $C_\mathrm{d}^*$ and $C_\mathrm{d}^{**}$ for  $\alpha=45^{\circ}$, different domain sizes, and different boundary conditions}
\label{P4-fig:P4-Cl-diffA}
\end{figure}

To assess the variation of the force coefficients with $A$ and BC, we started with simulations for all three BCs with $A=500c$ after doing careful grid independence checks similar to those described in detail by \citet{golmirzaee2024some}.  The number of cells and execution time 
for different values of $A$ and the three BCs
are shown in Table \ref{P4-table:DomTime-45}.
%%%%%%%%%%%%%%%%%%%%%%%%%%
\begin{table}
\centering
\caption{Comparison of the execution time and number of cells for BC-3 and PVSBC for different domain sizes.}
\begin{tabular}{lcccc}
\hline
A/c & Number of cells & Time (BC-3) [s] & Time (PVBC) [s] & Time (PVSBC) [s] \\
\hline
$500$ & $9,576,744$ & $331,209$ & $421,489$ & $490,553$ \\
$100$ & $4,903,424$ & $58,826$ & $70,461$ & $97,594$ \\
$30$ & $2,551,808$ & $25,181$ & $48,280$ & $57,987$  \\
$10$ & $1,600,384$ & $15,705$ & $25,725$ & $34,999$ \\
\hline
\end{tabular}
\label{P4-table:DomTime-45}
\end{table}
%%%%%%%%%%%%%%%%%%%%%%%%%%
The PVSBC is about $50\%$ slower than BC-3 at $A=500c$ and over twice as slow as BC-3 at $A=10c$.  It would need to have a significant advantage in accuracy at any $A$ to be more attractive.
Table \ref{P4-table:ClCd-Dom-45} presents the force and moment coefficients  for BC-3, PVBC, and PVSBC, at different $A$. The reason for adding the superscripts ``*'' and ``**''  will be described  below.  Clearly, there is close agreement between all three BCs when $A=500c$. This  allows us to use these results to infer the accuracy for other domain sizes.  The data for $C_\mathrm{l}$ is plotted in Fig. \ref{P4-fig:P4-Cl-diffA}. The most surprising aspect is the near-coincidence of the BC-3 and PVBC results.  
For  $\alpha \le 14^{\circ}$, the PVBC improves the accuracy of $C_\mathrm{d}$ and $C_\mathrm{l}$ compared to the BC-3 \cite{golmirzaee2024some}; however, once the drag becomes significant,  the influence of the PS condition becomes more critical than that of the PV. The other surprising feature is that the variation of any coefficient when normalized by the value at $A=500c$ is also very similar. This similarity is obvious from Table \ref{P4-table:ClCd-Dom-45} so we omitted  $C_\mathrm{d}$ and $C_\mathrm{m}$ from Fig. \ref{P4-fig:P4-Cl-diffA} in the interests of brevity.  The similar trends of the three coefficients may be associated with the dominance of the pressure in causing both forces and moment which, for a thin airfoil, will result in a net force in the direction normal to the chord.  On the other hand, the close agreement between BC-3 and PVBC suggests another cause.

%%%%%%%%%%%%%%%%%%%%%%%%%%
\begin{table}
\centering
\addtolength{\tabcolsep}{-0.3em}
\caption{Comparison of the force and moment coefficients for BC-3, PVBC, PVSBC, and different domain sizes.}
\begin{tabular}{lcccccccccccc}
\hline
 &&  & BC-3 &  &&  & PVBC & &&  & PVSBC & \\
\cmidrule{3-5} \cmidrule{7-9}
\cmidrule{11-13}
$A/c$ && $C^{**}_\mathrm{d}$ & $C^{**}_\mathrm{l}$ & $C^{**}_\mathrm{m}$ && $C^{**}_\mathrm{d}$ & $C^{**}_\mathrm{l}$ & $C^{**}_\mathrm{m}$ && $C^{*}_\mathrm{d}$ & $C^{*}_\mathrm{l}$ & $C^{*}_\mathrm{m}$ \\
\hline
$500$ && $0.8978$ & $0.9098$ & $-0.2206$ & & $0.8978$ & 
$0.9099$ & 
$-0.2207$ & & $0.8973$ & $0.9094$ & $-0.2205$ \\
$100$ && $0.9000$ & $0.9121$ & $-0.2212$ & & $0.9001$ & $0.9122$ & $-0.2212$ & & $0.8977$ & $0.9098$ & $-0.2207$ \\
$30$ && $0.9082$ & $0.9204$ & $-0.2232$ & & $0.9085$ & $0.9208$ & $-0.2234$ & & $0.9002$ & $0.9124$ & $-0.2213$ \\
$10$ && $0.9417$ & $0.9546$ & $-0.2316$  & & $0.9423$ & $0.9553$ & $-0.2319$ & & $0.9152$ & $0.9276$ & $-0.2250$ \\
\hline
\end{tabular}
\label{P4-table:ClCd-Dom-45}
\end{table}
%%%%%%%%%%%%%%%%%%%%%%%%%%
%%%%%%%%%%%%%%%%%%%%%%%%%%
BC-3 and PVBC  prevent any mass or momentum flux across the T and B boundaries, as do fixed, slip, symmetry or periodic BCs at T and B.  In these cases, Eq. (\ref{P4-eq:Continuity-3}) can be written as
%%%%%%%%%%%%%%%%%%%%%%%%%%
\begin{equation}
4u_{**} A + \int_O u_{\mathrm{v}} \mathrm{d}y = 0,  
\label{P4-eq:conv-usuv}
\end{equation}
%%%%%%%%%%%%%%%%%%%%%%%%%%
where $u_{**}$ is the average velocity exiting I and O. The integral in Eq. (\ref{P4-eq:conv-usuv}) is fixed by the drag, and so $u_{**}$ can be significant.  This is possibly the most important result of our investigation.  It means, for example, the average inlet velocity, $U_{\mathrm{I}}$, which is a user-input to any CFD simulation,  must now be identified as being less than $U_\infty$ when $D$ is sufficiently high and $A$ sufficiently small.  We have
%%%%%%%%%%%%%%%%%%%%%%%%%%
\begin{equation}
\frac{D}{\rho} = -U_\infty \int_O u_{\mathrm{v}} \mathrm{d}y \approx -(U_{\mathrm{I}} +u_{**})\int_O u_{\mathrm{v}} \mathrm{d}y.
\label{P4-eq:impulse-usuv}
\end{equation}
%%%%%%%%%%%%%%%%%%%%%%%%%%
Combining Eqs. (\ref{P4-eq:conv-usuv}) and (\ref{P4-eq:impulse-usuv}) and using the drag coefficient based on $U_{\mathrm{I}}$, denoted as $C^{**}_{\mathrm{d}}$, which is determined by the simulation, results in
%%%%%%%%%%%%%%%%%%%%%%%%%%
\begin{equation}
\frac{u_{**}}{U_{\mathrm{I}}}=\frac{1}{2}\left( 
\sqrt{1+\frac{C^{**}_{\mathrm{d}}c}{2A}}-1
\right).
\label{P4-eq:u**}
\end{equation}
%%%%%%%%%%%%%%%%%%%%%%%%%%
Then, the corrected drag coefficient can be found using
%%%%%%%%%%%%%%%%%%%%%%%%%%
\begin{equation}
\frac{C^*_{\mathrm{d}}}{C^{**}_{\mathrm{d}}}
= \left( 
1+ \frac{u_{**}}{U_{I}} \right)^{-2}.
\label{P4-cd**}
\end{equation}
%%%%%%%%%%%%%%%%%%%%%%%%%%
The superscript ``*'' is used for the PVSBC results in Fig. \ref{P4-fig:P4-Cl-diffA},  Table \ref{P4-table:ClCd-Dom-45}, and Eq. (\ref{P4-cd**}) for two reasons.  First, the PVSBC results should not need the correction just described.  There is, however, a further correction for all BCs that follows from \cite{imai1951}, and will be described below.  In other words, adding $u_{**}$ brings $U_{\mathrm{I}}$ closer to $U_\infty$, but does not make them equal. 

Equation (\ref{P4-cd**}) can be applied also to $C^{**}_{\mathrm{l}}$ and $C^{**}_{\mathrm{m}}$. We propose to call this the ``Lagally-Filon'' (LF) correction to honour these pioneering scientists whose work has been  neglected. The LF correction to $C^{**}_{\mathrm{d}}$ is shown in Fig. \ref{P4-fig:P4-Cl-diffA} and brings the BC-3 and PVBC results much closer to those for the PVSBC. 

\citet{rainbird2015blockage} applied a wind tunnel blockage correction given by their Eq. (A2) which can be written as
%%%%%%%%%%%%%%%%%%%%%%%%%%
\begin{equation}
\frac{C^*_{\mathrm{d}}}{C^{**}_{\mathrm{d}}}=\frac{1-e_{sc}}{(1+e_{sc}+e_{wc})^2},
\label{P4-ap2}
\end{equation}
%%%%%%%%%%%%%%%%%%%%%%%%%%
where $e_{wc}= C_\mathrm{d}^{**}c/(8A)$ and $e_{sc}=\zeta c^2/(4A^2)$ where $\zeta = 0.237 \pi^2/48$ and the constant $0.237$ applies to the NACA 0012 airfoil.
Figure \ref{P4-fig:P4-Cl-diffA} also shows this correction which is close to Eq. (\ref{P4-cd**}).  Applying the PVSBC, however, gives $C_\mathrm{d}$ at any $c/A$ closer to the correct values.  Further, the LF and wind tunnel corrections are not large enough by a factor of around two when $c/A \approx 0.1$.  

\begin{figure}
\centering
\includegraphics[width=1\linewidth]{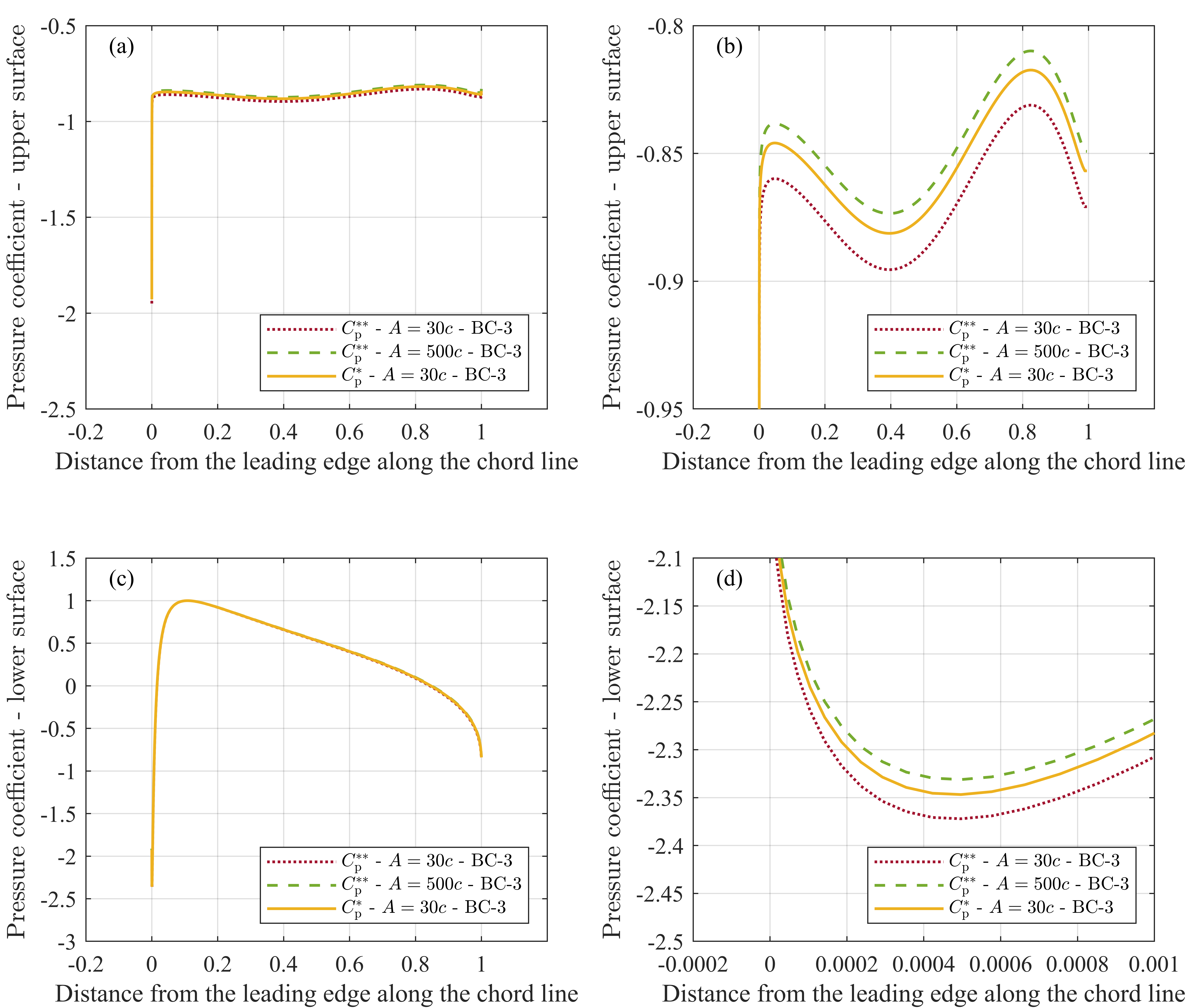}
\caption{Comparison of the corrected and uncorrected surface pressure distribution for BC-3 and $A=30c$ with that of $A=500c$. $C^{*}_{\mathrm{p}}$ is determined from $C^{**}_{\mathrm{p}}$ by Eq. (\ref{P4-eq:Cp*})}
\label{P4-fig:Cp}
\end{figure}

The pressure coefficient,
%%%%%%%%%%%%%%%%%%%%%%%%%%
\begin{equation}
C_{\mathrm{p}}^{**}=\frac{p-p_I}{\frac{1}{2}\rho U_{\mathrm{I}}^2},
\end{equation}
%%%%%%%%%%%%%%%%%%%%%%%%%%
where $p_I$ is the static pressure at the inlet, can also be easily corrected using Bernoulli's equation to write $p_I$ in terms of $p_\infty$ and Eq. (\ref{P4-eq:u**})
%%%%%%%%%%%%%%%%%%%%%%%%%%
\begin{equation}
C^*_{\mathrm{p}}=\frac{p-p_{\infty}}{\frac{1}{2}\rho (U_{\mathrm{I}} + u_{**})^2}=\left(
1-\frac{u_{**}}{U_{\infty}}
\right)^2\left( C_{\mathrm{p}}^{**}
- 1\right)
+1,
\label{P4-eq:Cp*}
\end{equation}
%%%%%%%%%%%%%%%%%%%%%%%%%%
so the correction makes no difference at the forward stagnation point but elsewhere, the corrected $C_{\mathrm{p}}^{*}$ values for BC-3 and $A=30c$ are closer to those for $A=500c$, Fig. \ref{P4-fig:Cp}. Part (a) of the figure shows the surface pressure is approximately constant due to the separated flow covering the entire upper surface.
%%%%%%%%%%%%%%%%%%%%%%%%%%
%%%%%%%%%%%%%%%%%%%%%%%%%%
\subsection{Comparison with previous numerical and experimental results}
\begin{figure}
\centering
\includegraphics[width=0.7\linewidth]{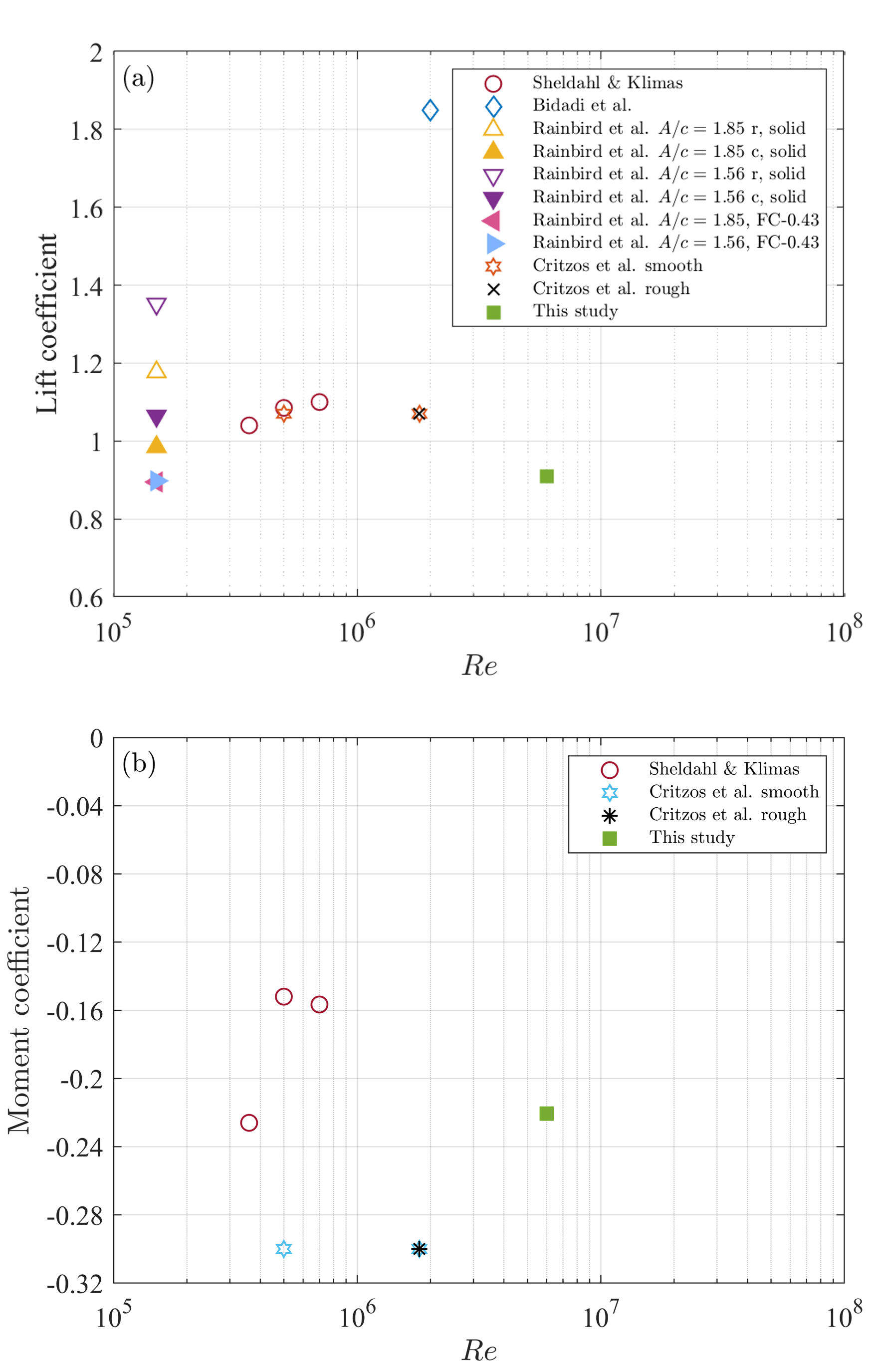}
\caption{Measurements and calculations of the lift and moment coefficients. The values for ``This study" are taken from $A=500c$ results.}
\label{P4-fig:Experiments}
\end{figure}

One reason for choosing  the NACA 0012 airfoil for our simulations is the large amount of data for it, taken over a wide range of $Re$. Figure \ref{P4-fig:Experiments} shows the data we found for $Re > 10^5$. All except that of \citet{bidadi2023mesh}, are experimental. Our primary purpose is not to assess the turbulence model, but there is value in judging the impact of BCs on the agreement between simulation and measurement.
%%%%%%%%%%%%%%%%%%%%%%%%%%
%%%%%%%%%%%%%%%%%%%%%%%%%%

\citet{sheldahl1981aerodynamic} produced the experimental data at the closest $Re$ ($0.7\times10^6$) to our simulations. Note that this reference provides tabulated data up to $Re = 10^7$, but all data above $Re = 0.7\times10^6$ was extrapolated from the measured data.  As in most experiments, the airfoil model was mounted in the middle of the tunnel; the distance from the model to the sidewalls gives $A/c=7$.  When its boundary layers are ignored, a sidewall causes the same blockage as BC-3 and PVBC, so we apply LF correction, Eq. (\ref{P4-cd**}), to their values of  $C^{**}_\mathrm{d}$ and $C^{**}_\mathrm{l}$ of $1.135$ and $1.1$, respectively.  This gives $C^{*}_\mathrm{d} =1.091$ and $C^{*}_\mathrm{l} =1.058$. 
Judging by Fig. \ref{P4-fig:P4-Cl-diffA}, a further correction of similar magnitude is necessary.

\citet{chris1995aerodynamic} used similar values of $A$, whereas \citet{rainbird2015blockage} used a closed-section wind tunnel with high blockage, 
$A/c = 1/(2\times0.27)$ and $1/(2\times0.32)$,
and a ``tolerant'' tunnel with slatted sidewalls to allow expansion of the flow around the airfoil.  The tolerant tunnel  values of $C_{\mathrm{d}}$ and  $C_{\mathrm{l}}$  are close to our results.  The measured and corrected closed-section data, denoted by subscripts ``r'' and ``c'', respectively, are shown in Fig. \ref{P4-fig:Experiments}.

The only simulation considered here was by \citet{bidadi2023mesh} who used the $k-\omega$ SST model along with more sophisticated ones that solved unsteady equations. They examined the effect of mesh resolution in both the wall-normal and spanwise directions. For  $\alpha=45^{\circ}$, the $k-\omega$ SST model significantly overestimated $C_\mathrm{l}$ and $C_\mathrm{d}$; however, the detached eddy model IDDES, with a minimum spanwise resolution of $24$ cells per chord length, more accurately predicted the aerodynamic forces.

It is reasonable to conclude that using the PVSBC produces RANS results for forces and moments in this highly turbulent and separated flow that are within $10\%$ of the experimental values once a full correction (the LF or wind tunnel correction plus an additional correction of similar magnitude according to fig. \ref{P4-fig:P4-Cl-diffA})  is applied to the experimental data.
We note, however, that
when $D$ becomes large enough and $A$ small enough, the direct effect of the BC-3 and PVBC must be felt by the flow over the body and the LF correction will not allow accurate extrapolation of the results to infinite $A$.  In other words, the LF correction is an incomplete correction, as demonstrated in Fig. \ref{P4-fig:P4-Cl-diffA}.

%%%%%%%%%%%%%%%%%%%%%%%%%%
\subsection{Velocity over the domain}
\label{P4-subsec:velprevor}
\begin{figure}
\centering
\includegraphics[width=0.9\linewidth]{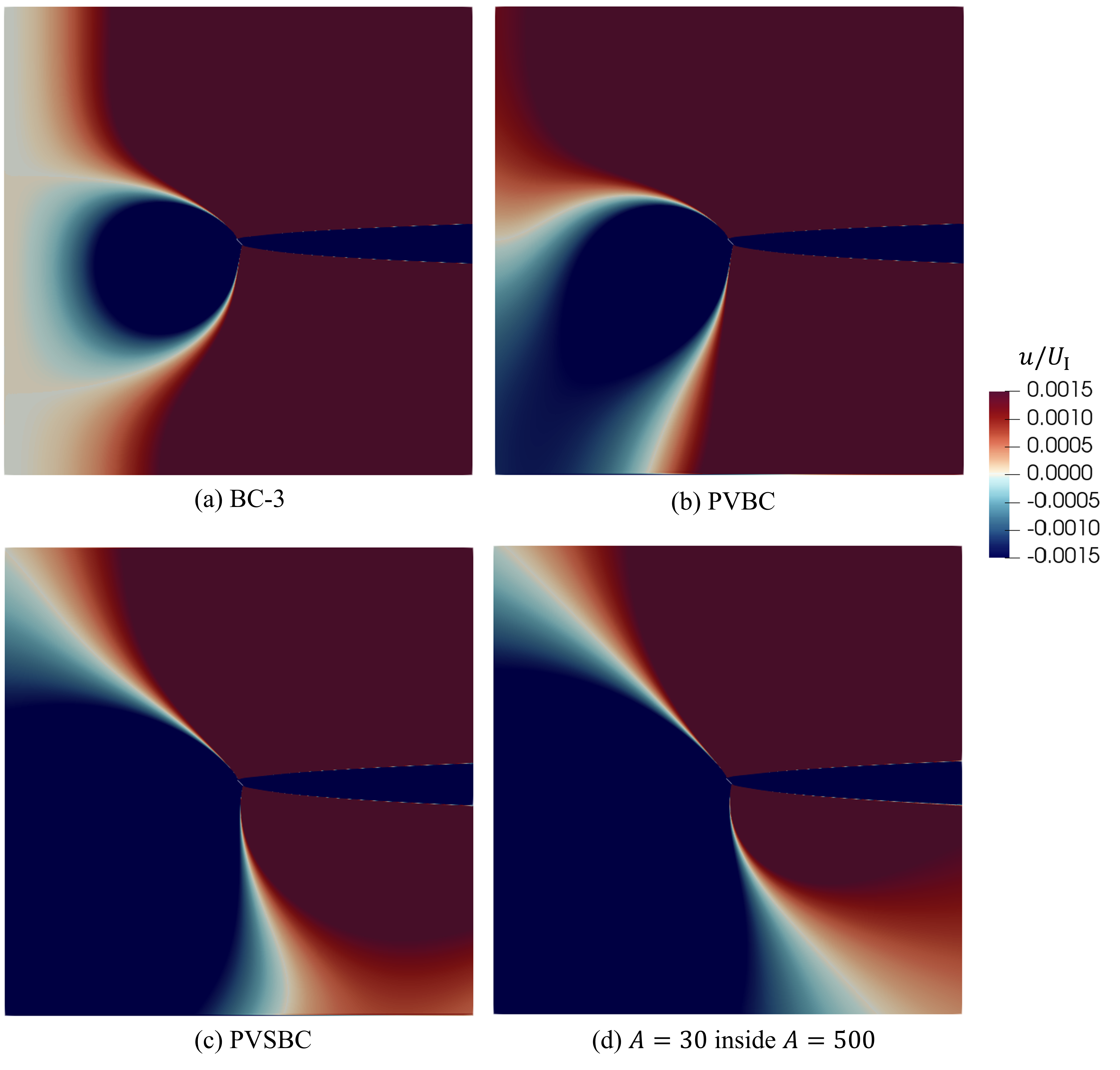}
\caption{$u$ over the computational domain for $A=30c$ and the BCs indicated.  Part (d) shows the same domain taken from $A=500c$ with the PVSBC.}
\label{P4-fig:U-V-velocity}
\end{figure}

Parts (a), (b), and (c) of Fig. \ref{P4-fig:U-V-velocity} show $u$  contours over the CD for $A=30c$ and the BCs indicated.  For comparison,  the contours in (d) are for $A=500c$ and the PVSBC within a square of $A=30c$. Clearly, $u$ from the PVSBC is closer to the ``correct'' distribution in part (d) and this is true also for $v$ which is not shown in the interests of brevity.

%%%%%%%%%%%%%%%%%%%%%%%%%%
%%%%%%%%%%%%%%%%%%%%%%%%%%
\begin{figure}
\centering
\includegraphics[width=0.9\linewidth]{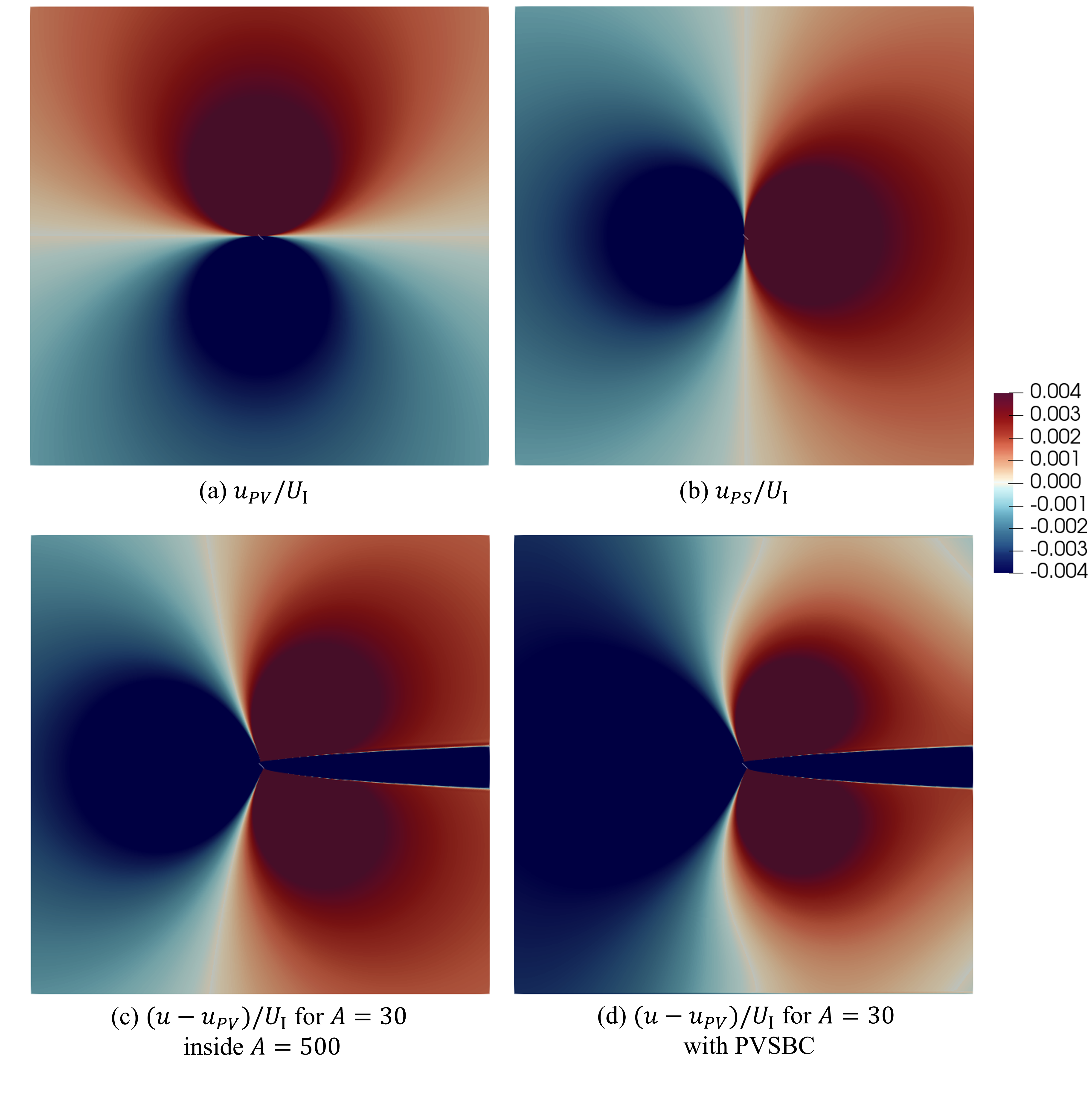}
\caption{Contribution of point source and vortex to $u$ for $A=30c$ and PVSBC.  Part (c) shows the same domain taken from $A=500c$ and PVSBC.}
\label{P4-fig:u-v-PV-PS}
\end{figure}
%%%%%%%%%%%%%%%%%%%%%%%%%%
\begin{figure}
\centering
\includegraphics[width=1\linewidth]{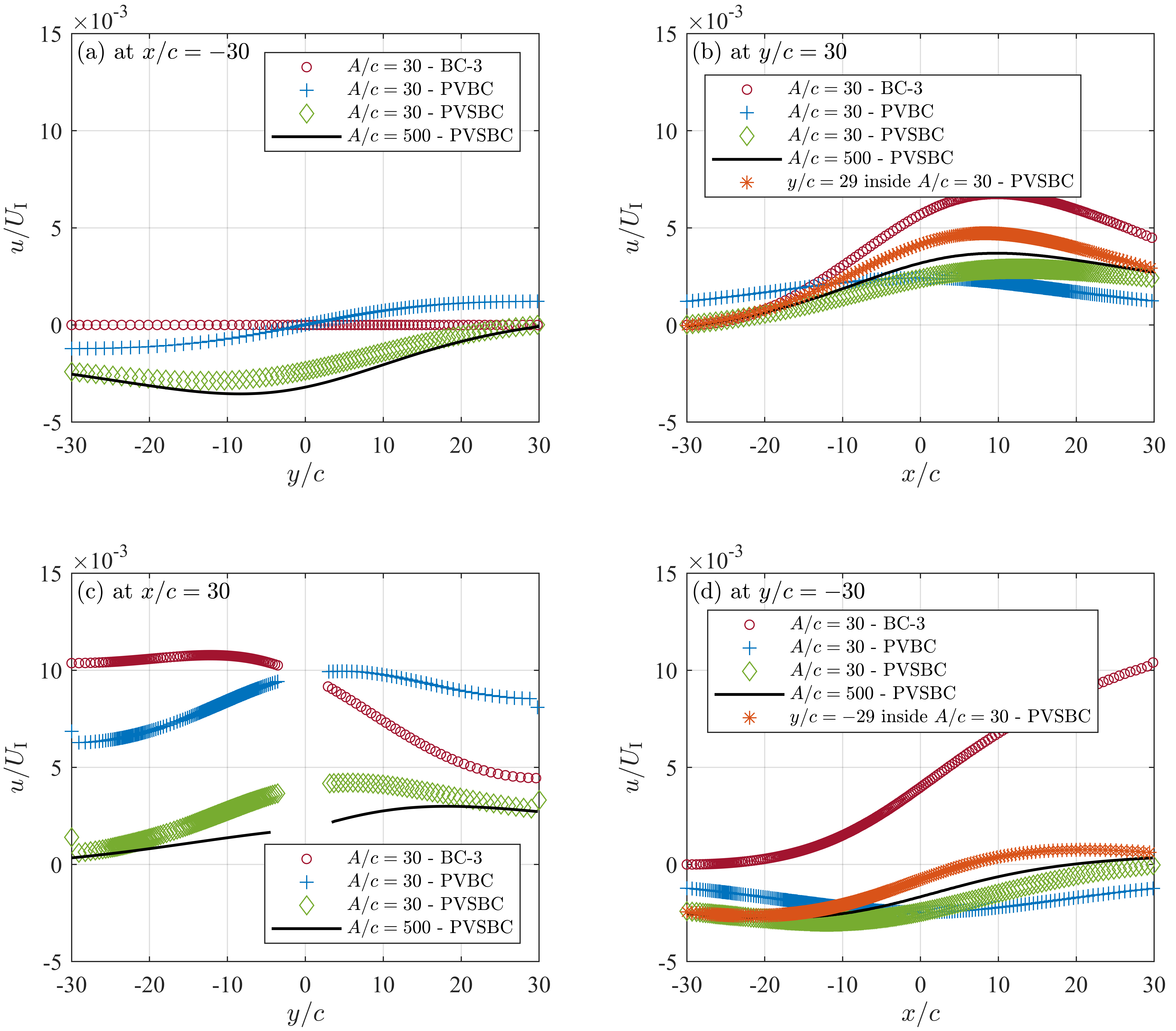}
\caption{Comparison of $u/U_{\mathrm{I}}$ with different boundary conditions at the four boundaries. Only every tenth point is  shown for clarity.}
\label{P4-fig:u-BC-3-PVBC-PVSBC}
\end{figure}
%%%%%%%%%%%%%%%%%%%%%%%%%%
\begin{figure}
\centering
\includegraphics[width=1\linewidth]{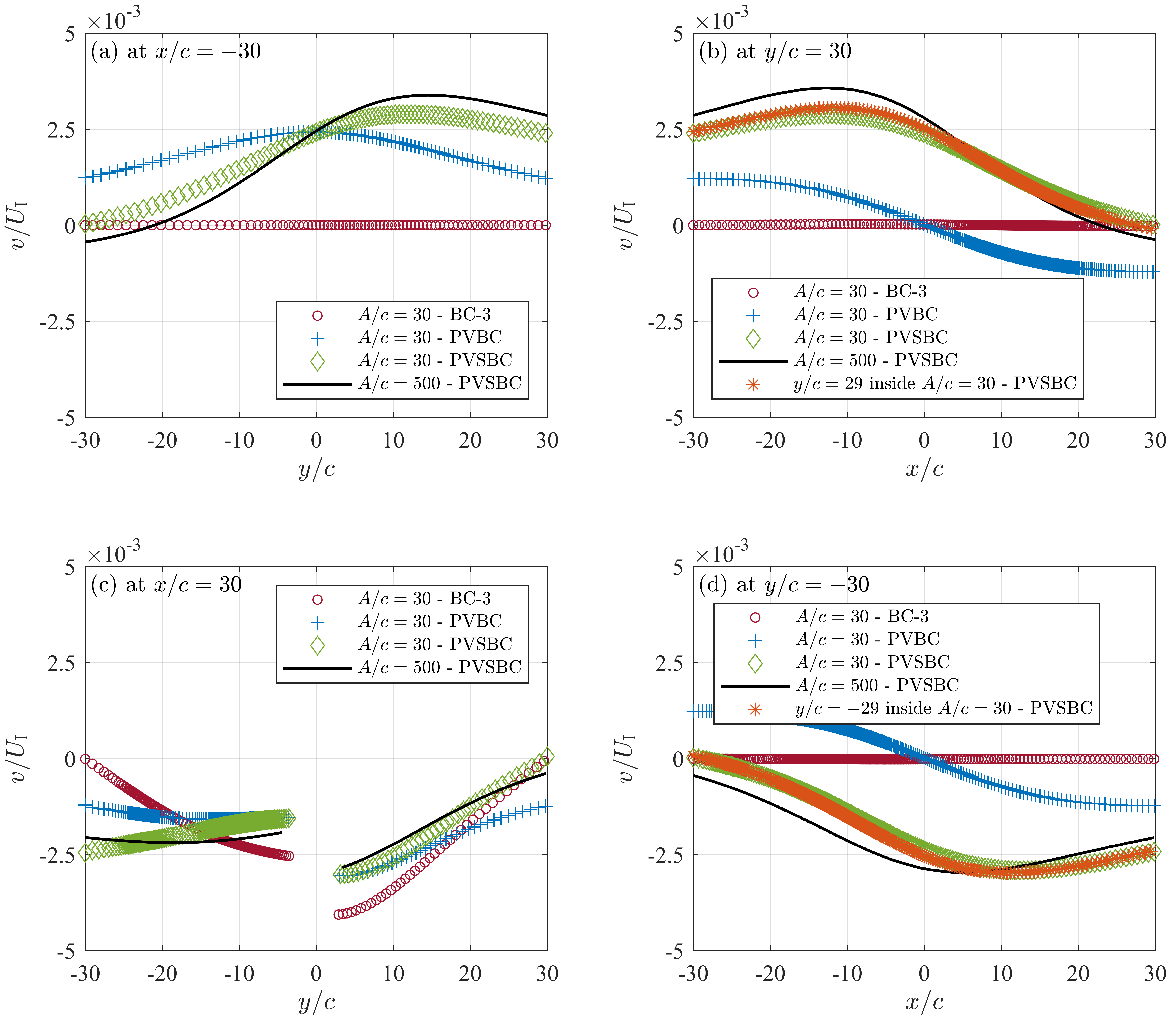}
\caption{Comparison of $v/U_{\mathrm{I}}$ with different boundary conditions at different boundaries. Only every tenth point is  shown for clarity.}
\label{P4-fig:v-BC-3-PVBC-PVSBC}
\end{figure}
%%%%%%%%%%%%%%%%%%%%%%%%%%

Parts (a) and (b) of Fig. \ref{P4-fig:u-v-PV-PS} show the contours of the PV and PS contributions to $u$, from Eq. (\ref{P4-eq:PVPS-velocity}).  Part (c)  was obtained by subtracting the PV contribution from the total $u$ over a domain of $A=30c$ situated within $A=500c$ using the PVSBC and (d) plots  the same for the actual $A=30c$ domain. Again the similarity between (c) and (d) is strong but it is clear that the wake has complicated the induced velocity field, especially for $x >0$.  Further evidence for this is the profiles of $u$ and $v$ along the boundaries in Figs. \ref{P4-fig:u-BC-3-PVBC-PVSBC} and  \ref{P4-fig:v-BC-3-PVBC-PVSBC},  respectively, for $A=30c$ and the three BCs tested.  The vorticity-carrying $u_{\mathrm{v}}$ and $v_{\mathrm{v}}$ in the wake are not plotted.  The black lines in both figures are the $A=500c$ results for the PVSBC. Starting with the former figure: part (a) shows the impact of the inlet BC on the induced mass flow: it is zero relative to $U_{\mathrm{I}}$ for BC-3 and PVBC and clearly negative for the PVSBC and the ``correct'' black distribution.  Similarly, at the outlet, part (c) shows the higher exit flow for BC-3 and PVBC.  The average of the outflow $u$ should be $2u_{**}$ for both these BCs and this is shown to be approximately the case in Table \ref{P4-Table:Avgu-2u**}.
  
\begin{figure}
\centering
\includegraphics[width=0.9\linewidth]{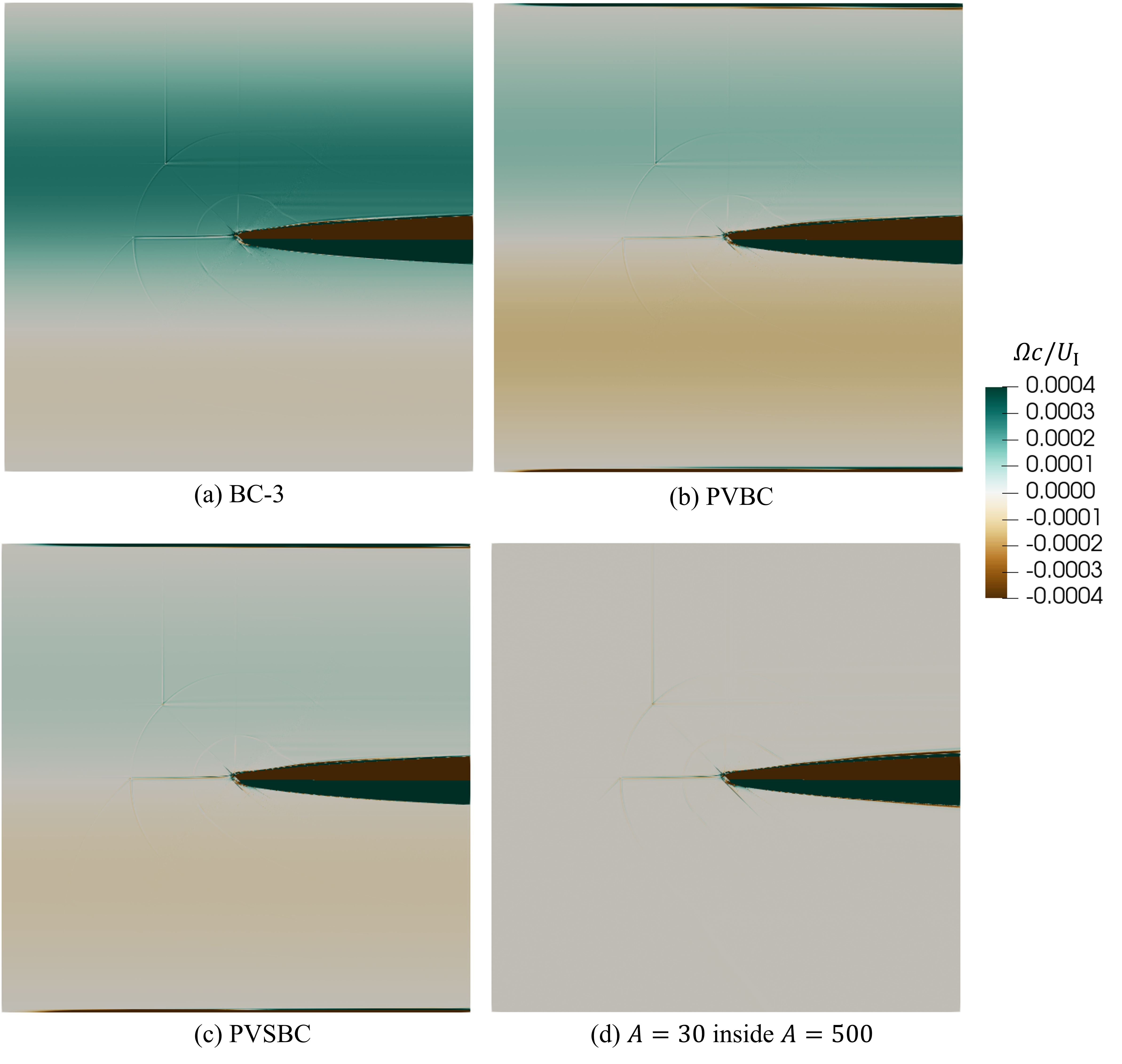}
\caption{$\Omega$ over the computational domain for $A=30c$ and the BCs indicated.  Part (d) shows the same domain taken from $A=500c$ with the PVSBC.}
\label{P4-fig:Omega-BCs}
\end{figure}

%%%%%%%%%%%%%%%%%%%%%%%%%%
\begin{table}
\centering
\caption{Comparison of average $u/U_{\mathrm{I}}$  outside the wake and $u_{**}/U_{\mathrm{I}}$ from Eq. (\ref{P4-eq:u**}) for BC-3 and PVBC.}
\begin{tabular}{lcccc}
\hline
$A/c$ & $2u_{**}/U_{\mathrm{I}}$ (BC-3) & $u/U_{\mathrm{I}}$ (BC-3) &
$2u_{**}/U_{\mathrm{I}}$ (PVBC) & $u/U_{\mathrm{I}}$ (PVBC)
\\
\hline
$500$  & $0.00045$ & $0.00046$ & $0.00045$ & $0.00046$ \\
$100$ & $0.0022$ & $0.0024$ & $0.0022$ & $0.0024$ \\
$30$ & $0.0075$ & $0.0084$ 
& $0.0075$ & $0.0084$ \\
$10$ & $0.0233$ & $0.0281$ 
& $0.0233$ & $0.0285$ \\
\hline
\end{tabular}
\label{P4-Table:Avgu-2u**}
\end{table}
%%%%%%%%%%%%%%%%%%%%%%%%%%
A close inspection of the outlet $u$ reveals a jump in $u$ as O meets T and B for both the PVBC and PVSBC.  This is associated with the formation of vorticity layers along the top and bottom sidewalls, Fig. \ref{P4-fig:Omega-BCs}, that were observed at lower $\alpha$ by \citet{golmirzaee2024some}. We argue below that these layers are the mechanism by which that BC satisfies the moment equation in the form where all vorticity exits in the wake by reducing the effective size of the CD. 
%%%%%%%%%%%%%%%%%%%%%%%%%%
%%%%%%%%%%%%%%%%%%%%%%%%%%
Parts (b) and (d) of Fig. \ref{P4-fig:u-BC-3-PVBC-PVSBC} show the distribution of $u$ for the PVSBC and $A=30c$ along $y=\pm 29c$, that is, sufficiently far into  the domain to exclude the vorticity layers. The effective change in the T and B boundaries changes the contributions to the moment equation as will be discussed in Section \ref{P4-subsec:forcemomentA30PVSBC}.  Overall, $u$ and $v$ for the PVSBC follow the correct results reasonably closely and so only small differences are associated with the BC not being consistent with the moment equation.  The PVSBC slightly under-estimates the magnitude of the inlet mass flux, and over-estimates the magnitude of the outlet flux.  This behavior is the basis for the second, Imai correction to $U_{\mathrm{I}}$ that is discussed below.

%%%%%%%%%%%%%%%%%%%%%%%%%%
\subsection{Force and moment balances for $A=500c$ and PVSBC}
\label{P4-subsec:ForceMomentA500}
To test the impulse equations for $L$, $D$, and $M$, the PVSBC domain for $A=500c$ was subdivided into square domains corresponding to $A = 30c, 100c, $ and $300c$ and the impulse equations of Section \ref{P4-sec:ImpulseEq} applied. Table \ref{P4-table:D-A500PVSBC} shows the terms in Eq. (\ref{P4-eq:Fx-impulse-Uinf-reduced-2}) for the drag. The impulse equation is accurate and the second order terms decay in the far-field.  The decay, however, is slower than for the similar terms in the lift equation (\ref{P4-eq:Fy-impulse-Uinf-4}) shown in Table \ref{P4-table:L-A500PVSBC}. It is clear that  the quadratic terms for $D$ do not sum to zero for any $A$, and the same is true for $L$, at least for $A=30c$. The non-zero sum in all cases is due largely to the outlet terms and may be due partly to the application of the Neumann BC.  There may also be a contribution from the interaction between the vorticity field of the wake and the inviscid flow, as evidenced by the difference in $v$ across the wake evident in Fig. \ref{P4-fig:v-BC-3-PVBC-PVSBC}(c) which is due to the growth of the wake, but is not included in the point source and vortex model.
%%%%%%%%%%%%%%%%%%%%%%%%%%
\begin{table}
\centering
\caption{Drag balance. The table entries show contributions to $C_{\mathrm{d}}$. From Table \ref{P4-table:ClCd-Dom-45}, $C_{\mathrm{d}}$ = 0.8973}
\begin{tabular}{lccc}
\hline
Term & $A=30c$ & $A=100c$ & $A=300c$ \\
\hline
$\int_{I} (u^2/2-v^2/2)\mathrm{d}y$ & $7.5 \times 10^{-5}$ & $6\times 10^{-6}$ & $-7\times 10^{-7}
$
\\
$\int_{T} uv\mathrm{d}x$ & $-4.6 \times 10^{-4}
$ & $-1.2\times 10^{-4}$ & $-3.9\times 10^{-5}$
\\
$-\int_{O} y\Omega \mathrm{d}y$ 
& $1.0216$ & $0.9600$ & $0.9277$
\\
$-\int_{O} yu\Omega \mathrm{d}y$ 
& $-0.0621$ & $-0.0315$ & $-0.0179$
\\
$-\int_{O} (u^2/2-v^2/2)\mathrm{d}y$
& $-0.0623$ & $-0.0316$ & $-0.0179$
\\
$\int_{B} uv\mathrm{d}x$ & $3.1 \times 10^{-4}$ & $9.7\times 10^{-5}$ & $3.1\times 10^{-5}$
\\
Sum & $0.8971$ & $0.8969$ & $0.8919$ \\
Percentage Error&  $0.02$ & $0.04$  & $0.61$ \\
\hline
\end{tabular}
\label{P4-table:D-A500PVSBC}
\end{table}
%%%%%%%%%%%%%%%%%%%%%%%%%%
\begin{table}
\centering
\caption{Lift balance.  The table entries show contributions to $C_{\mathrm{l}}$.   From Table \ref{P4-table:ClCd-Dom-45}, $C_{\mathrm{l}}$ = 0.9094.  }
\begin{tabular}{lccc}
\hline
Term & $A=30c$ & $A=100c$ & $A=300c$ \\
\hline
$U_{\infty} \Gamma$ & $0.9114$ & $0.9098$ & $0.9085$ \\
$\int_{I} uv\mathrm{d}y$ & $-4.3 \times 10^{-4}$ & $-1.1 \times 10^{-4}$ & $-3.4 \times 10^{-5}$ \\
$\int_{T} (u^2/2-v^2/2)\mathrm{d}x$ & $4.6 \times 10^{-5}$ & $9.0 \times 10^{-6}$ & $3.9 \times 10^{-6}$ \\
$-\int_{O} uv \mathrm{d}y$ & $-0.0018$ & $-5.4 \times 10^{-4}$ & $-1.8 \times 10^{-4}$ \\
$-\int_{B} (u^2/2-v^2/2)\mathrm{d}x$ & $0.0001$ & $1.6 \times 10^{-5}$ & $3.5 \times 10^{-6}$ \\
Sum & $0.9093$ & $0.9092$ & $0.9083$ \\
Percentage Error & $0.006 $ & $0.023 $ &  $0.12 $ \\
\hline
\end{tabular}
\label{P4-table:L-A500PVSBC}
\end{table}
%%%%%%%%%%%%%%%%%%%%%%%%%%
 
Table \ref{P4-Table:Moments} shows the moment balance is the least accurate, presumably because the appearance of $x$ and $y$ in the integrands emphasizes small errors.  Further, it is clear that the quadratic terms in rows 4 to 7 decay, but the one involving $\Omega$ in row 6 is still of magnitude of $30\%$ of $M$ in the $A=300c$ column.  As explained above, the leading order vorticity integral grows monotonically.

%%%%%%%%%%%%%%%%%%%%%%%%%%
\begin{table}
\centering
\caption{Moment balance.  The table entries show contributions to $C_{\mathrm{m}}$. From Table \ref{P4-table:ClCd-Dom-45}, $C_{\mathrm{m}} = -0.2205$.}
\begin{tabular}{lccc}
\hline
 Term & $A=30c$ & $A=100c$ & $A=300c$ \\
\hline 
$U_{\infty}(\int_{I}
uy \mathrm{d}y
-\int_{O} uy \mathrm{d}y
-\int_{T} vy \mathrm{d}x
+\int_{B} vy \mathrm{d}x)$ &
$-0.1174$ & $-0.1166$ & $-0.1725$ \\
 $U_{\infty}(-\int_{I} vx \mathrm{d}y
+\int_{O} vx \mathrm{d}y
-\int_{T} ux \mathrm{d}x
+\int_{B} ux \mathrm{d}x)$ &
$0.2142$ & $0.2881$ & $0.4127$ \\
$-U_{\infty}\int_{O}
\Omega y^2 \mathrm{d}y/2$ & 
$-0.3510$ & $-0.4131$ & $-0.4782$ \\
$\int_{I}
(-uvx-v^2 y/2 +u^2y/2)\mathrm{d}y$ & 
$-0.0207$ & $-0.0179$ & $-0.0164$ \\
$\int_{T}
(-uvy
+v^2 x/2
-u^2x/2)\mathrm{d}x$ & 
$-0.0227$ & $-0.0194$ & $-0.0182$ \\
$\int_{O}
(-y^2u\Omega/2
+uvx
+v^2y/2-u^2y/2)\mathrm{d}y$ & 
$0.0914$ & $0.0753$ & $0.0653$ \\
$\int_{B}
(uvy-v^2 x/2+u^2x/2) \mathrm{d}x$ &
$-0.0143$ & $-0.0152$ & $-0.0150$ \\
Sum &
$-0.2205$ & $-0.2188$ & $-0.2223$ \\
%\hline
Percentage Error & 
$0$ & $0.77$ & $0.82$ \\
\hline
\end{tabular}
\label{P4-Table:Moments}
\end{table}
%%%%%%%%%%%%%%%%%%%%%%%%%%
\subsection{Force and moment balances for $A=30c$ and PVSBC}
\label{P4-subsec:forcemomentA30PVSBC}
Comparing the impulse equation and simulation results for $A=30c$ with PVSBC gives $0.02\%$ and $0.82\%$ percentage error for the drag and lift, respectively.

The second column in Table \ref{P4-Table:Moments-A30-PVSBC} shows the moment balance for a CV moved inward by distance $c$  from T and B, while the I and O boundaries  were kept at $x=\pm 30c$.  
The moment balance has $0.05\%$ error for this CV, but for the CV coincident with the CD, column 3, the error is about $66\%$. There are significant differences in the term in the second row resulting from the differences in $u$ shown in Fig. \ref{P4-fig:u-BC-3-PVBC-PVSBC}(b) and (d).  The reason for the difference is the development of vorticity layers along T and B (see Fig. \ref{P4-fig:Omega-BCs}) which occurred for the PVBC and the PVSBC, and which the results in Table \ref{P4-Table:Moments-A30-PVSBC} suggest are the mechanism by which the PVSBC becomes consistent with the moment equation. We did not investigate the vorticity layers in detail as this would, presumably, have required considering the equivalent $y^+$ and reducing the cell size near T and B to improve resolution in the boundary layers.
%%%%%%%%%%%%%%%%%%%%%%%%%%
\begin{table}
\centering
\caption{Moment balance for $x= \pm 30c$ and $(i)$ $y=\pm 29c$ and $(ii)$ $y=\pm 30c$ inside $A=30c$ with PVSBC. The table entries show contributions to $C_{\mathrm{m}}$. From Table \ref{P4-table:ClCd-Dom-45}, $C_{\mathrm{m}}=-0.2213$. } 
\begin{tabular}{lcc}
\hline
Term & $y=\pm 29c$ & $y=\pm 30c$ \\
\hline
$U_{\infty}(\int_{I}
uy \mathrm{d}y
-\int_{O} uy \mathrm{d}y
-\int_{T} vy \mathrm{d}x
+\int_{B} vy \mathrm{d}x)$ & 
$-0.2264$ & $-0.2286$
\\
$U_{\infty}(-\int_{I} vx \mathrm{d}y
+\int_{O} vx \mathrm{d}y
-\int_{T} ux \mathrm{d}x
+\int_{B} ux \mathrm{d}x)$ & 
$0.3214$ & $0.1785$
\\
$-U_{\infty}\int_{\mathrm{wake}}
\Omega y^2 \mathrm{d}y/2$ & 
$-0.3482$ & $-0.3482$
\\
$\int_{I}
(-uvx-v^2 y/2 +u^2y/2)\mathrm{d}y$ & 
$-0.0159$ & $-0.0162$
\\
$\int_{T}
(-uvy
+v^2 x/2
-u^2x/2)\mathrm{d}x$ & 
$-0.0257$ & $-0.0162$
\\
$-\int_{\mathrm{wake}}
y^2u\Omega/2
\mathrm{d}y$ & 
$0.0206$ & $0.0206$
\\
$\int_{O}
[uvx
+v^2y/2-u^2y/2]\mathrm{d}y$ & 
$0.0599$ & $0.0594$
\\
$\int_{B}
[uvy-v^2 x/2+u^2x/2] \mathrm{d}x$ & 
$-0.0069$ & $-0.0165$
\\
Sum & 
$-0.2212$ & $-0.3672$
\\
%\hline
Percentage Error & 
$0.05$ & $65.9$
\\
\hline
\end{tabular}
\label{P4-Table:Moments-A30-PVSBC}
\end{table}
%%%%%%%%%%%%%%%%%%%%%%%%%%
\subsection{Development of the wake}
The development and deflection of the wake is an important feature of the moment balance and, presumably, the accuracy of the BCs. Further, the symmetry of the velocity profile influences the logarithmic divergence of the vorticity integral in the moment equation and  the turbulence Reynolds number of the wake must be known to use Imai's equation for the far-field streamfunction as discussed in the next Section.  

Self-preservation of wakes with the invariant drag integral (\ref{P4-cd1a}) leads to the scaling of the wake thickness, $\delta \sim (x-x_0)^{1/2}$ and the minimum velocity as  \text{$u_\mathrm{m} \sim (x-x_0)^{-1/2}$}, where $x_0$ is the effective origin of the fully-developed flow, e.g.,  \cite{townsend1976}.  This scaling is shown in Fig. \ref{P4-fig:P4-xu2wakewidth}, for  $A=500c$ and PVSBC. The most obvious feature is the high degree of symmetry of the wake, showing that the increase in the vorticity integral in Table \ref{P4-table:y2Omega} is dominated by the wake deflection. It is remarkable that the angular momentum flux exiting the CD through the wake is greater than the airfoil moment for all values of $A$ that were investigated.  Goldstein's simple model for the logarithmic growth of the integral involving $C_\mathrm{d}C_\mathrm{l}$ in the last column is accurate to within a constant, suggesting a non-zero effective $y$-direction origin for the wake.

\begin{figure}
\centering
\includegraphics[width=1\linewidth]{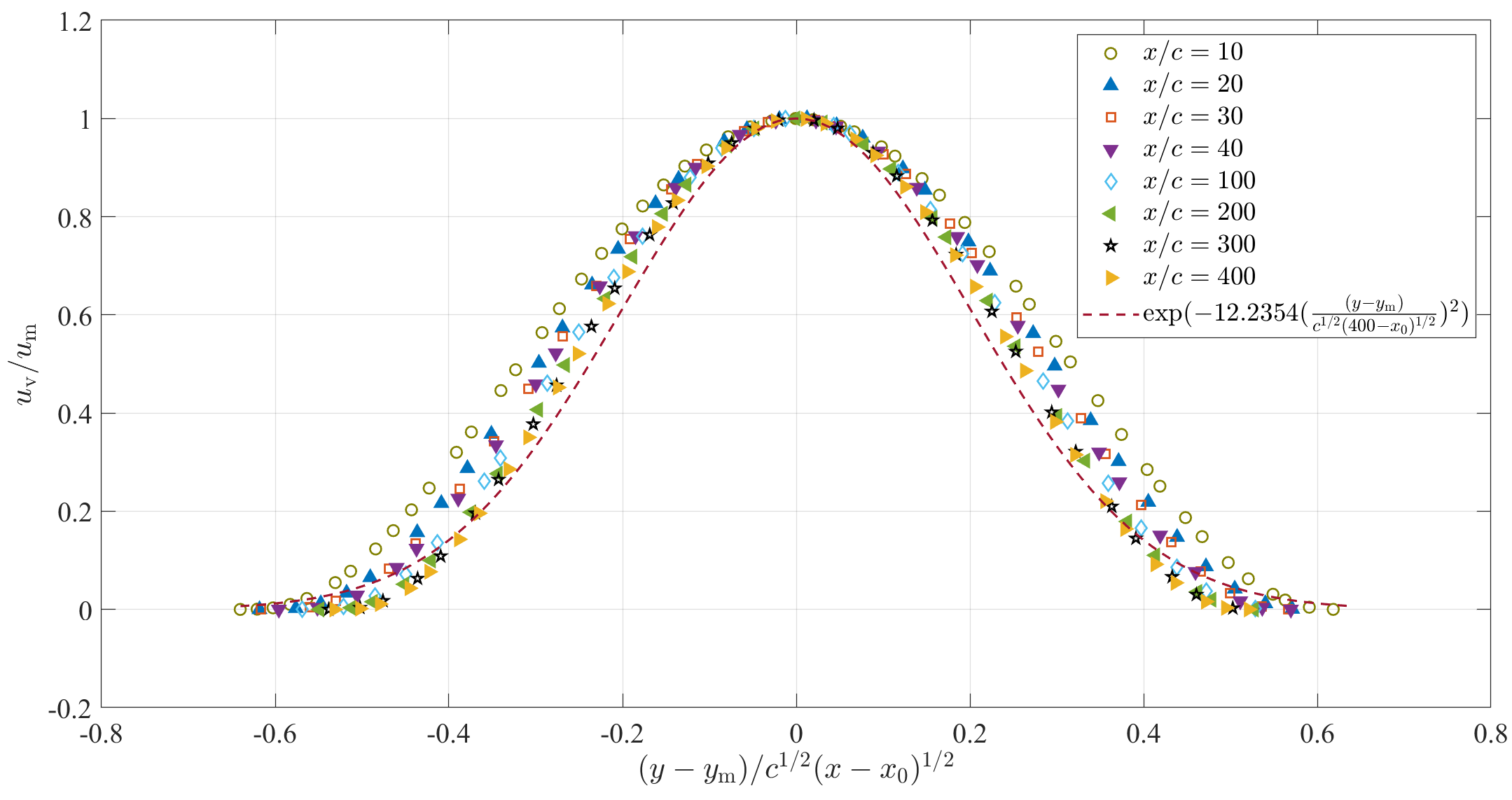}
\caption{Velocity profile in the wake in terms of $u_{\mathrm{v}}/u_\mathrm{m}$ vs $(y-y_\mathrm{m})/c^{1/2}(x-x_0)^{1/2}$. Only every fifth point from the simulations is plotted for clarity.}
\label{P4-fig:uvum-x0-A500-PSBC}
\end{figure}

\begin{figure}
\centering
\includegraphics[width=0.75\linewidth]{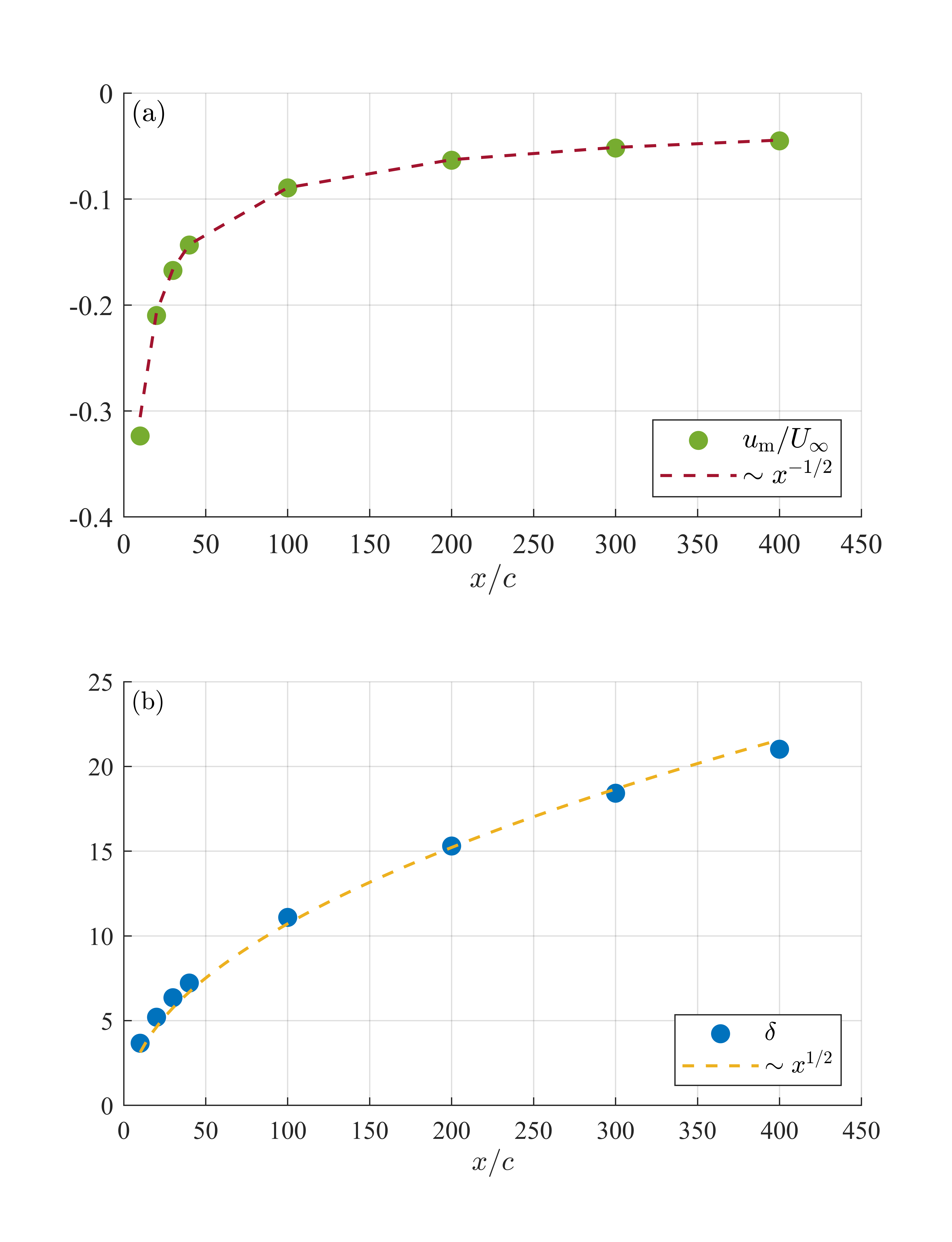}
\caption{Development of the minimum velocity, $u_\mathrm{m}$, in part (a) and the wake thickness, $\delta$, in part (b).  The lines of fit are: $u_\mathrm{m}/U_{\infty}=-0.885c^{1/2}(x-1.622)^{-1/2}$ and $\delta=1.081c^{1/2}(x-1.622)^{1/2}$}
\label{P4-fig:P4-xu2wakewidth}
\end{figure}

The velocity profile in a laminar wake in the far-field is given on page 349 of \cite{batchelor2000}
%%%%%%%%%%%%%%%%%%%%%%%%%%
\begin{equation}
    \frac{u_{\mathrm{v}}}{U_\infty}=\frac{u_{\mathrm{v}}}{u_{\mathrm{m}}}\frac{u_{\mathrm{m}}}{U_\infty}= -\frac{C_{\mathrm{d}}Re^{1/2} c^{1/2}}{4\sqrt{\pi}(x-x_0)^{1/2}}\exp\left(-\frac{Re(y-y_{\mathrm{m}})^2}{4 c (x-x_0)}\right),
    \label{P4-lamwake}
\end{equation}
%%%%%%%%%%%%%%%%%%%%%%%%%%
where $y_{\mathrm{m}}$ is the location of the minimum velocity, $u_{\mathrm{m}}$, in the wake and the exponential term gives $u_{\mathrm{v}}/u_{\mathrm{m}}$. For a turbulent wake with a constant eddy viscosity, $Re$ can be interpreted as the turbulence Reynolds number, in which the eddy viscosity replaces the molecular viscosity. Figure \ref{P4-fig:uvum-x0-A500-PSBC} shows the velocity in the wake, and Fig. \ref{P4-fig:P4-xu2wakewidth}(a) the streamwise development of $u_{\mathrm{m}}$ in terms of Eq. (\ref{P4-lamwake}) with the turbulence $Re =48$.  $x$ must exceed $100c$ 
for the magnitude of $u_{\mathrm{m}}/U_{\infty}$ to be less than $0.1$.   Part (b) of Fig. \ref{P4-fig:P4-xu2wakewidth} depicts the streamwise growth of $\delta$ defined as the distance between the two points where $\Omega=0$
is first encountered on either side of $y_{\mathrm{m}}$.  Finally, the accuracy of the Neumann BC in the wake was assessed by determining the root mean square of $(\partial v/\partial x)/\Omega_{\max}$. At $x=10c$, the value was $0.0056$, at $x=30c$, it was $0.0011$, and at $x=300c$, $0.0001$.
%%%%%%%%%%%%%%%%%%%%%%%%%%
\begin{table}
\centering
\caption{The development of the vorticity integral in the moment equation}
\begin{tabular}{lcccc}
\hline
$x/c$ & $-(2/U_{\infty}c^2)\int_{\mathrm{wake}}
\Omega y^2 \mathrm{d}y/2$ & $y_\mathrm{m}/c$ & $2y_\mathrm{m} D/\rho U^2_{\infty} c^2$ & $-C_\mathrm{d}C_\mathrm{l}\ln(x/c)/4\pi$
\\
\hline
$10$ & $-0.3078$ & $-0.2593$ & $-0.2327$ & $-0.1495$ \\
$20$ & $-0.3275$ & $-0.3056$ & $-0.2743$ & $-0.1945$ \\
$30$ & $-0.3473$ & $-0.3293$ & $-0.2955$ & $-0.2209$ \\
$40$ & $-0.3609$ & $-0.3497$ & $-0.3138$ & $-0.2395$ \\
$100$ & $-0.4055$ & $-0.4146$ & $-0.3720$ & $-0.2990$ \\
$200$ & $-0.4412$ & $-0.4638$ & $-0.4162$ & $-0.3440$ \\
$300$ & $-0.4642$ & $-0.4937$ & $-0.4430$ & $-0.3704$ \\
$400$ & $-0.4829$ & $-0.5163$ & $-0.4633$ & $-0.3890$ \\
\hline
\end{tabular}
\label{P4-table:y2Omega}
\end{table}
%%%%%%%%%%%%%%%%%%%%%%%%%%
%%%%%%%%%%%%%%%%%%%%%%%%%%
%%%%%%%%%%%%%%%%%%%%%%%%%%
\section{The Imai correction}
\label{P4-sec:ImaiCorr}
\begin{figure}
\centering
\includegraphics[width=0.75\linewidth]{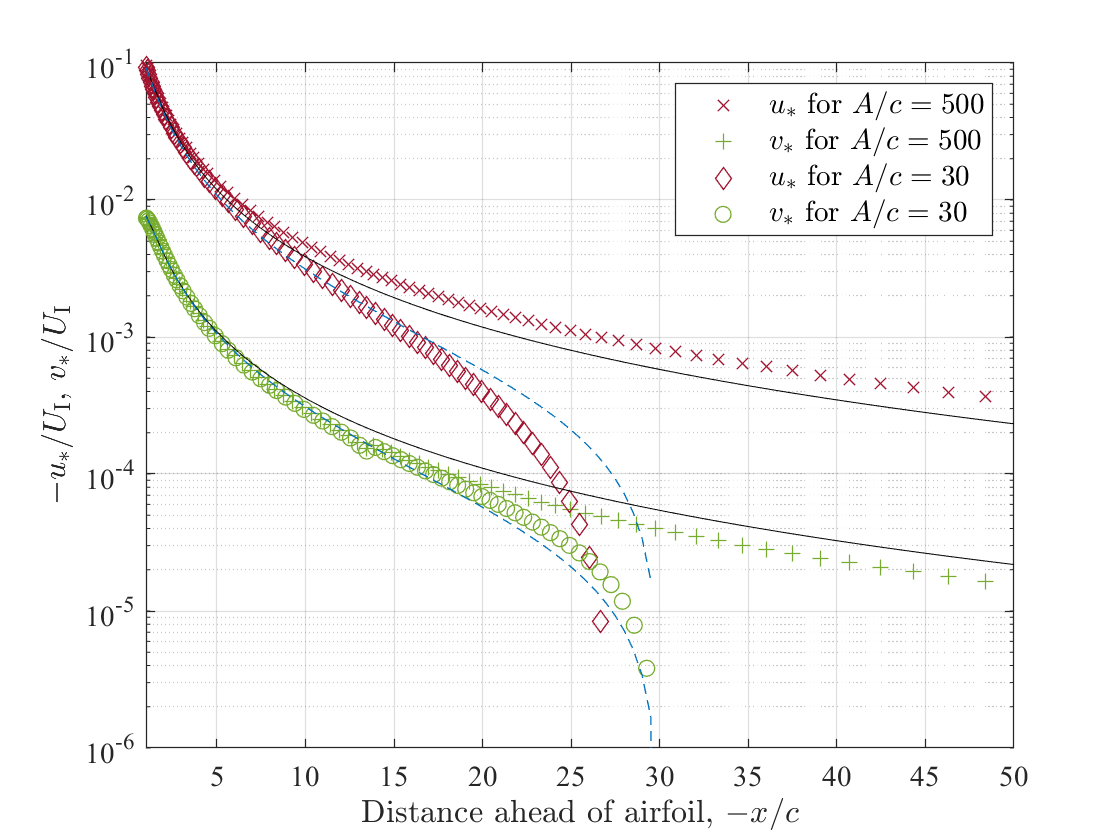}
\caption{Variation of $u_*(x,0)$ and $v_*(x,0)$ ahead of the airfoil for $A =30c$ and $500c$. The solid and dashed lines show the least square fit to Eq. (\ref{P4-imai2}) for $A = 500c$ and $30c$, respectively. Only every tenth point from the simulations is plotted.}
\label{P4-fig:uv}
\end{figure}

The LF correction to $U_{\mathrm{I}}$ gives $U_{\infty} \approx U_{\mathrm{I}}+u_{**}$ for BCs that cause sidewall blockage.  Figure \ref{P4-fig:u-BC-3-PVBC-PVSBC}, however, suggests that the PVSBC is not sufficient to make $u$ at the inlet and outlet coincident with the far-field distributions shown as the solid lines.  This shows a lingering difference between $U_{\mathrm{I}}+u_{**}$ and $U_\infty$ for this BC, as implied by  the PVSBC being inconsistent with the logarithmic divergence of the vorticity integral in the moment equation.  \citet{imai1951} addressed the divergence for a deflected laminar wake behind a lifting body. His analysis should be applicable in the present case because the turbulent wake is well-described by a constant eddy viscosity.  The far-field streamfunction is his Eq. (11.6) which involves, in the present case, the turbulence Reynolds number.  Unfortunately, there appears to be an error in this equation as it does not force the $x$-axis to be a streamline for a symmetric body with $\Gamma=0$.  Nevertheless, the form of the departure of $(u,v)$ from the PVSBC values, which we denote as $(u_*,v_*)$, along the $x$-axis suggests that
%%%%%%%%%%%%%%%%%%%%%%%%%%
\begin{equation}
\frac{u_*}{U_{\mathrm{I}}}(x,y=0)=\frac{a_* c^2}{x^2}+\frac{b_* c^2 \ln(x^2/c^2)}{x^2}-\frac{a_* c^2}{A^2}-\frac{b_* c^2 \ln(A^2/c^2)}{A^2}
\label{P4-imai2}
\end{equation}
%%%%%%%%%%%%%%%%%%%%%%%%%%
with a similar equation for $v_*$. 
Equation (\ref{P4-imai2}) forces the BCs $u_*(A)=v_*(A) =0$.
Figure \ref{P4-fig:uv} shows the effect of the BC  at $A=30c$ on $u_{*}$ propagates farther downstream than for $v_{*}$.  This difference also occurs in normal flow impingement on a solid surface where the region over which the streamwise velocity decreases to satisfy the inviscid BC of zero normal velocity,  is generally larger than the viscous Hiemenz region due to the no-slip condition on the tangential velocity.  In the present case, the magnitude of $v_*$ is apparently sufficiently small not to cause a boundary layer to form at the inlet.  Figure \ref{P4-fig:uv} shows the least squares fit of Eq. (\ref{P4-imai2}) to the results for $A=30c$ and $500c$.  
%%%%%%%%%%%%%%%%%%%%%%%%%%
%%%%%%%%%%%%%%%%%%%%%%%%%%
For $u_*$, $(a_*,b_*)=(0.0928, 0.0635)$ for $A=500c$ and $(0.0926, 0.0598)$ for $A=30c$. For   $v_*$ and $A=500c$, $(a_*,b_*)= (0.0077, 0.0061)$, and $(0.0076, 0.0063)$ for $A=30c$. It appears, therefore, that the data for $A=30c$ can be used to estimate $u_*/U_{\mathrm{I}}$ at $x=A$ from Eq. (\ref{P4-imai2}) by allowing $A \to \infty$.  This gives $u_*/U_{\mathrm{I}} = 5.5 \times 10^{-4}$ for $A=30c$ which is too small to justify correcting any of our results and is small compared to $u_{**}$ in Table \ref{P4-Table:Avgu-2u**}. \citet{imai1951} also gave an equation for the moment -- his Eq. (14.8) -- which we have not pursued as it is possibly in error.

The small magnitude of the Imai correction suggests that it is not worthwhile to pursue modifications to the PVSBC to make it consistent with the moment equation.  Figures \ref{P4-fig:u-BC-3-PVBC-PVSBC} and \ref{P4-fig:v-BC-3-PVBC-PVSBC} imply that the changes would be small, and they must leave $\Gamma$ and $\Lambda$ unaltered as well as ensuring that the quadratic terms in the lift and drag equation continue to sum to zero for any $A$. The possible exception to this statement could be flows where the moment is critical, such as for vertical-axis turbines.
%%%%%%%%%%%%%%%%%%%%%%%%%%
\section{Summary and conclusion}
\label{P4-sec:SummaryConclusion}
We considered the simulation of steady, two-dimensional, incompressible flow over a NACA 0012 airfoil at a high Reynolds number, $6\times10^6$, using the Spalart-Allmaras turbulence model.  By reducing the numerical and other sources of error in the simulations, we were able to document the changes in lift, drag, and moment with changes in the boundary conditions (BCs).  The 45$^\circ$ angle of attack was high enough for the lift and drag to be comparable and the moment about the quarter-chord to be significant.   From our earlier study of the same airfoil at lower angles \cite{golmirzaee2024some}, we tested a set of commonly-used boundary conditions (BC-3) and our version of a point vortex boundary condition (PVBC) which \citet{golmirzaee2024some} found to give the most accurate results for a given domain size and, therefore, to be faster to execute for a given level of accuracy. The strength of the point vortex was chosen to satisfy the Kutta-Joukowsky equation which we derived from the impulse equation for lift using a control volume (CV) coincident with the square computational domain. The corresponding equation for the drag is the Lagally-Filon equation relating drag to source strength.

Our results show that when the drag is significant, the PVBC performs no better than BC-3 because both  prevent  transfer of mass or momentum through the top and bottom sidewalls.  A point source must be added to produce the PVSBC, the only BC of the three we tested, that is consistent with the Kutta-Joukowsky  and   the Lagally-Filon equations. The hypothesis presented in the Introduction, that maximizing consistency of the boundary conditions will maximize accuracy for a given domain size, has been shown to be correct.  Nevertheless, the practical situation is complicated by the unexpectedly larger execution time required by the PVSBC and the success of the Lagally-Filon correction in substantially reducing the error in the lift, drag, and moment of the less consistent, but faster executing, boundary conditions.

Blockage by the top and bottom sidewalls makes the user-specified inlet velocity smaller than the freestream velocity at infinity. Somewhat surprisingly, the difference in these velocities is easily calculated by what  we proposed to call the ``Lagally-Filon'' correction derived in 
Section \ref{P4-sec:results}.  Applied to the BC-3 and PVBC results for a range of domain sizes, the correction makes the lift, drag and moment coefficients close to the PVSBC values which required considerably more execution time. The correction to the airfoil surface pressure distribution was also derived and shown to improve the results for a BC causing blockage.  Computational sidewall blockage is similar in principle to sidewall blockage in experiments on airfoils and other bodies. We show that the Lagally-Filon correction is very close to the blockage correction used by \citet{rainbird2015blockage} for their experiments on the NACA 0012 at a Reynolds number of 150,000. The measured lift and drag coefficients from their ``blockage tolerant'' wind tunnel are very close to the PVSBC values computed here.  When the measurements of \citet{sheldahl1981aerodynamic} and \citet{chris1995aerodynamic} at Reynolds numbers closer to ours are corrected, and the correction is approximately doubled to be in line with the large $c/A$ results in fig. \ref{P4-fig:P4-Cl-diffA}, they are about 10\% larger than our PVSBC results.

The PVSBC is still not consistent with the impulse form of the moment equation because its vorticity integral is logarithmically divergent in distance downstream of the airfoil whereas its boundary integral terms are zero for the PVSBC wherever the point source and vortex are placed within the CV. This problem was  found by \citet{filon1928}, explained as a consequence of the deflection of the wake in the direction opposite to the lift by \citet{goldstein1933}, and addressed analytically by \citet{imai1951} for a laminar wake.  This restriction to laminar flow is probably not significant as the present turbulent wake is well-described by a constant eddy viscosity. 
Unfortunately, however, there appears to be an error in Imai's analysis which we did not resolve, and  it cannot be directly applied. We showed, however, that a simple equation for the velocities ahead of the airfoil can be used to assess the magnitude of the correction for wake deflection, which we called the ``Imai'' correction. For the present simulations, the Imai correction was too small to justify its application.
 
There are three main reasons to believe that the inconsistency between the PVSBC and the moment equation is not significant for the present flow.  The first is the very small level of the residual streamwise velocity along the $y$-axis, due to the higher-order effects of wake deflection and found by subtracting the point source velocity (the contribution from the point vortex is zero along the $x$-axis).  In other words, the Imai correction had a negligible impact on the lift, drag, and moment. Secondly, the fractional change in the moment coefficient with domain size was very similar to that of the lift and drag coefficients, for any BC.  Thirdly, whereas the PVSBC generates vorticity layers adjacent to the top and bottom walls of the computational domain, probably as a consequence of the inconsistency, Table \ref{P4-Table:Moments-A30-PVSBC} shows these layers to be  thin and the domain within them to accurately obey the moment equation in the form that assumes all vorticity leaves through the outlet.  Nevertheless, it remains to be seen whether these reasons apply also to simulations where the moment is critical, such as vertical-axis turbines. In this regard, the remarkable symmetry of the present wake about its point of minimum velocity, shown in Fig. \ref{P4-fig:uvum-x0-A500-PSBC}, may be significant because the logarithmic divergence could change with substantial asymmetry and Imai's moment equation includes a term dependent on the asymmetry.  Finding the correct version of Imai's streamfunction and moment equations should be very valuable.

%%%%%%%%%%%%%%%%%%%%%%%%%%
\bmhead{Acknowledgments}
This work is supported by NSERC Discovery Grant \mbox{RGPIN\slash04886-2017} and the Schulich endowment to the University of Calgary.
%%%%%%%%%%%%%%%%%%%%%%%%%%
\section*{Declarations}

\begin{itemize}
\item Funding: 
NSERC Discovery Grant \mbox{RGPIN\slash04886-2017} and the Schulich endowment to the University of Calgary.
\item Competing interests:
The authors declare that they have no competing inter-
ests.
\item Availability of data and materials:
The code used to implement the PVBC and PVSBC and the data produced in this study are available from the corresponding author upon reasonable request.
\item Authors' contributions:
Narges Golmirzaee: Conceptualization, Data curation, Formal Analysis, Investigation, Methodology, Software, Validation, Visualization, Writing – original draft, Writing – review \& editing. 
David Wood: Conceptualization, Data curation, Formal Analysis, Funding acquisition, Investigation, Methodology, Project administration, Supervision, Validation, Visualization, Writing – original draft, Writing – review \& editing.
\end{itemize}
%%%%%%%%%%%%%%%%%%%%%%%%%%

\end{document}